\pdfoutput=1
\documentclass[useAMS,usenatbib]{mn2e}
\usepackage{amssymb}
\usepackage{bbm}
\usepackage{txfonts}
\usepackage{psfrag}
\usepackage{graphicx}
\usepackage{natbib}
\usepackage{if then}
\usepackage{longtable}

\def\del#1{{}}

\title{Bayesian non-linear large scale structure inference of the Sloan Digital Sky Survey data release 7}
\author[Jens Jasche \\$^{1}$, Francisco Shu Kitaura, Cheng Li, Torsten A. En\ss lin]
       {Jens Jasche $^{1}$, Francisco S. Kitaura $^{2}$, Cheng Li $^{1}$  ,Torsten A. En\ss lin $^{1}$ \\$^{1}$ Max-Planck-Institut f\"{u}r Astrophysik , Karl-Schwarzschild Stra\ss e 1,  D-85748 Garching, Germany\\$^{2}$ SNS, Scuola Normale Superiore, Piazza dei Cavalieri 7, 56126 Pisa, Italy}
\begin{document}
\date{Submitted to MNRAS 12-Nov-2009}

\pagerange{\pageref{firstpage}--\pageref{lastpage}} \pubyear{2006}

\maketitle

\label{firstpage}

\begin{abstract}
In this work we present the first non-linear, non-Gaussian full Bayesian large scale structure analysis of the cosmic density field conducted so far. The density inference is based on the Sloan Digital Sky Survey data release 7, which covers the northern galactic cap. We employ a novel Bayesian sampling algorithm, which enables us to explore the extremely high dimensional non-Gaussian, non-linear log-normal Poissonian posterior of the three dimensional density field conditional on the data. These techniques are efficiently implemented in the HADES computer algorithm and permit the precise recovery of poorly sampled objects and non-linear density fields. The non-linear density inference is performed on a 750 Mpc cube with roughly 3 Mpc grid-resolution, while accounting for systematic effects, introduced by survey geometry and selection function of the SDSS, and the correct treatment of a Poissonian shot noise contribution. Our high resolution results represent remarkably well the cosmic web structure of the cosmic density field. Filaments, voids and clusters are clearly visible. Further, we also conduct a dynamical web classification, and estimated the web type posterior distribution conditional on the SDSS data. 
\end{abstract}


\section{Introduction}
Observations of the large scale structure have always attracted enormous interest, since they contain a wealth of information on the origin and evolution of our Universe. The details of structure formation are very complicated and involve many different physical disciplines ranging from quantum field theory, general relativity or modified gravity to the dynamics of collisionless matter and the behavior of the baryonic sector. Throughout cosmic history the interplay of these different physical phenomena therefore has left its imprints in the matter distribution surrounding us. Probes of the large scale structure, such as large galaxy surveys, hence enable us to test current physical and cosmological theories and will generally further our understanding of the Universe.

Especially a cosmographical description of the matter distribution will permit us to study details of structure formation mechanisms and the clustering behavior of galaxies as well as it will provide information on the initial fluctuations and large scale cosmic flows.
For this reason, several different methods to recover the three dimensional density or velocity field from galaxy observations have been developed and applied to existing galaxy surveys \citep[][]{1993PhRvE..47..704E,HOFFMAN1994,1994ASPC...67..171L,1994ApJ...423L..93L,1995A&AS..109...71Z,1995MNRAS.272..885F,1995ApJ...449..446Z,1997MNRAS.287..425W,1999ApJ...520..413Z,2001misk.conf..268V,2006MNRAS.373...45E,ERDOGDU2004,KITAURA2009B}. In particular, recently \citet{KITAURA2009} presented a high resolution three dimensional Wiener reconstruction of the Sloan Digital Sky Survey data release 6 data, which demonstrated the feasibility of high precision density field inference from galaxy redshift surveys.
These three dimensional density maps are interesting for a variety of different scientific applications, such as studying the dependence of galaxy properties on their cosmic environment, increasing the detectability of the integrated Sachs-Wolfe effect in the CMB or performing constrained simulations \citep[see e.g.][]{BISTOLAS1998,LEE2008,LEELI2008,FROMMERT2008,KLYPIN2003,LIBESKIND2009,2009MNRAS.397.2070M}. 

However, modern precision cosmology demands an increasing control of observational systematic and statistical uncertainties, and the means to propagate them to any finally inferred quantity in order not to draw wrong conclusion on the theoretical model to be tested.
For this reason, here we present the first application of the new Bayesian large scale structure inference computer algorithm HADES (HAmiltonian Density Estimation and Sampling) to data \citep[see][for a description of the algorithm]{JASCHE2009B}. 
HADES performs a full scale non-linear, non-Gaussian Markov Chain Monte Carlo analysis by drawing samples from the lognormal Poissonian posterior of the three dimensional density field conditional on the data. This extremely high dimensional posterior distribution, consisting of \(\sim 10^6\) or more free parameters, is explored via a numerically efficient Hamiltonian sampling scheme which suppresses the random walk behavior of conventional Metropolis Hastings algorithms by following persistent trajectories through the parameter space \citep[][]{DUANE1987,NEAL1993,NEAL1996}.
The advantages of this method are manyfold. Beside correcting observational systematics introduced by survey geometry and selection effects, the exact treatment of the non-Gaussian behavior and structure of the Poissonian shot noise contribution of discrete galaxy distributions, permits very accurate recovery of poorly sampled objects, such as voids and filaments. In addition, the lognormal prior has been demonstrated to be an adequate statistical description for the present density field and hence enables us to infer the cosmic density field deep into the non-linear regime \citep[see e.g.][]{HUBBLE1934,PEEBLES1980,COLES1991,GAZTANAGA1993,KAYO2001}. 
The important thing to remark about HADES is, that it does not only yield a single estimate, such as a mean, mode or variance, in fact it provides a sampled representation of the full non-Gaussian density posterior. This posterior encodes the full non-linear and non-Gaussian observational uncertainties, which can easily be propagated to any finally inferred quantity.

The application of HADES to Sloan Digital Sky Survey (SDSS) data therefore is the first non-linear, non-Gaussian full Bayesian large scale structure analysis conducted so far \citep[SDSS;][]{YORK2000}. In particular, we applied our method to the recent SDSS data release 7 (DR7) data \citep[DR7;][]{SDSS7}, and produced about 3TB of valuable scientific information in the form of \(40000\) high resolution non-linear density samples. The density inference is conducted on an equidistant cubic grid with side length \(750\) Mpc consisting of \(256^3\) volume elements. The recovered density field clearly reveals the cosmic web structure, consisting of voids, filaments and clusters, of the large scale structure surrounding us. 

These results provide the basis for forthcoming investigations on the clustering behavior of galaxies in relation to their large-scale environment. Such analyses yield valuable information about the formation and evolution of galaxies. In example, it has been known since long that physical properties such as morphological type, color, luminosity, spin parameter, star formation rate, concentration parameter, etc., are functions of the cosmic environment \citep[see e.g.][]{DRESSLER1980,POSTMAN1984,WHITMORE1993,LEWIS2002,GOMEZ2003,GOTO2003,ROJAS2005,KUEHN2005,BLANTON2005,BERNARDI2006,CHOI2007,PARK2007,LEE2008,LEELI2008}.

In this work we already conduct a preliminary examination of the dependence of stellar mass \(M_{\star}\) and \(g-r\) color of galaxies on their large-scale environment. However, more thorough investigations will be presented in following works.  
Analyzing galaxy properties in the large-scale environment also requires to classify the large scale structure into different cosmic web types. We do so by following the dynamic cosmic web type classification procedure as proposed by \citet{HAHN2007} with the extension of \citet{FORERO2009}. The application of this procedure to our results yields the cosmic web type posterior, which provides the probability of finding a certain web type (void, sheet, filament, halo) at a given position in the volume conditional on the SDSS data. This permits simple propagation of all observational uncertainties to the final analysis of galaxy properties.

The paper is structured as follows. We start by a brief review of the methodology in section \ref{Methodology}, particularly describing the lognormal Poissonian posterior and the Bayesian computer algorithm HADES. Additionally, here we describe the dynamic web classification procedure as mentioned above. In section \ref{DATA} we give a description of the SDSS DR7 data and present necessary data preparation steps required to apply the analysis procedure. Specifically, we describe the preparation of the linear observation response operator and the creation of the three dimensional data cube. 
In the following section \ref{RESULTS} we present the results obtained
from the non-linear, non-Gaussian sampling procedure. We start by
analyzing the convergence behavior of the Markov chain via a Gelman \&
Rubin diagnostic, followed by a discussion of the properties of
individual Hamiltonian samples. Here we also provide estimates for the
ensemble mean density field and according variance. These fields
depict remarkable well the properties of the cosmic web consisting of
voids, filaments and halos. Based on these results we perform a simple
cosmic web classification in section
\ref{web_classification_data}.
In section \ref{GALAXY_PROPERTIES}, we present a preliminary
examination on the correlation between the large-scale environment of
galaxies and their physical properties. In particular, here we study
the stellar mass and \(g-r\) color of galaxies in relation with the
density contrast \(\delta\). We conclude the paper in
section \ref{SUMMARYANDCONCLUSION} by summarizing and discussing the results.

\section{Methodology}
\label{Methodology}
In this section we give a brief review of the methods used for the large scale structure inference. In particular, we discuss the lognormal Poissonian posterior, and the according data model. Further, we give a description of the HADES algorithm and a dynamic cosmic web classification procedure.
\subsection{Lognormal Poissonian posterior}
\label{LOGNORMALPOISSONIANPOSTERIOR}
Precision inference of the large scale structure in the mildly and strongly non-linear regime requires detailed treatment of the non-Gaussian behavior of the large scale structure posterior. Although, the exact probability distribution for the density field in these regimes is not known, for a long time already it has been suggested that the fully evolved non-linear matter field can be well described by lognormal statistics \citep[see e.g.][]{HUBBLE1934,PEEBLES1980,COLES1991,GAZTANAGA1993,KAYO2001}. This phenomenological guess has been justified by the theoretical considerations of \citet{COLES1991}. They argue that assuming Gaussian initial conditions in the density and velocity distributions will lead to a log-normally distributed density field. It is a direct consequence of the continuity equation or the conservation of mass. In addition, the validity of the lognormal distribution as a description of the statistical properties of non-linear density fields has been evaluated in \citet{KAYO2001}.
In this work, they studied the probability distribution of
cosmological non-linear density fluctuations from N-body simulations
with Gaussian initial conditions. They found that the lognormal
distribution accurately describes the non-linear density field even up
to values of the density contrast of \(\delta \sim 100\). In addition,
recently \citet{KITAURA2009} analyzed the statistical properties of
the SDSS DR6 Wiener reconstructed density field, and confirmed a
lognormal behavior.

For all these reasons, we believe, that the statistical behavior of the non-linear density field can be well described by a multivariate lognormal distribution, as given by:
\begin{equation}
\label{eq:LogNormal_prior}
{\mathcal P}(\{s_k\}|Q)=\frac{1}{\sqrt{2\pi det(Q)}} e^{-\frac{1}{2}\sum_{ij} \left(ln(1+s_i)+\mu_i\right) Q^{-1}_{ij} \left(ln(1+s_j)+\mu_j\right)} \prod_k \frac{1}{1+s_k} \, ,
\end{equation}
where \(s_i\) is the density signal at the three dimensional cartesian position \(\vec{x}_i\), \(Q\) is the covariance matrix of the lognormal distribution and \(\mu_i\) describes a constant mean field given by:
\begin{equation}
\label{eq:MU}
\mu_i=\frac{1}{2}\sum_{l,m} Q_{lm} \, .
\end{equation}
This probability distribution, seems to be an adequate prior choice for reconstructing the present density field.

Studying the actual matter distribution of the Universe requires to draw inference from some observable tracer particle, such as a set of observed galaxies. Assuming galaxies to be discrete particles, their distribution can be described as a specific realization drawn from an inhomogeneous Poisson process \citep[see e.g.][]{LAYZER1956,PEEBLES1980,MARTINEZ2002}. The according probability distribution is given as:
\begin{equation}
\label{eq:Poissonian}
{\mathcal P}(\{N_k^{g}\}|\{\lambda_k\})= \prod_k \frac{{\left(\lambda_k\right)}^{N^{g}_k} e^{-\lambda_k}}{{N^{g}_k}!} \, ,
\end{equation}
where \(N_k^{g}\) is the observed galaxy number at position \(\vec{x}_k\) in the sky and  \(\lambda_k\) is the expected number of galaxies at this position.
The mean galaxy number is related to the signal \(s_k\) via:
\begin{equation}
\label{eq:data_model}
\lambda_k= R_k \bar{N}(1+B(s)_k)\, ,
\end{equation}
where \(R_k\) is a linear response operator, incorporating survey geometries and selection effects, \(\bar{N}\) is the mean number of galaxies in the volume and \(B(x)_k\) is a non-linear, non local, bias operator at position \(\vec{x}_k\).
The lognormal prior given in equation (\ref{eq:LogNormal_prior}) together with the Poissonian likelihood given in equation (\ref{eq:Poissonian}) yields the lognormal Poissonian posterior, for the density contrast \(s_k\) given some galaxy observations \(N_k^{g}\):
\begin{eqnarray}
\label{eq:LOGNORMALPOISSONIAN_POSTERIOR}
{\mathcal P}(\{s_k\}|\{N_k^{g}\}) &=& \frac{e^{-\frac{1}{2}\sum_{ij} \left(ln(1+s_i)+\mu_i\right) Q^{-1}_{ij} \left(ln(1+s_j)+\mu_j\right)}}{\sqrt{2\pi det(Q)}} \prod_l \frac{1}{1+s_l} \nonumber \\
& & \times \prod_k \frac{{\left(R_k \bar{N}(1+B(s)_k)\right)}^{N^{g}_k} e^{-R_k \bar{N}(1+B(s)_k)}}{{N^{g}_k}!}
\end{eqnarray}
It is important to note, that this is a highly non-Gaussian distribution, and non-linear reconstruction methods are required in order to perform accurate matter field reconstructions in the non-linear regime. In example, estimating the maximum a posteriori values from the lognormal Poissonian distribution involves the solution of implicit equations. Several attempts to use a lognormal Poissonian posterior for density inference have been presented in literature. These attempts date back to \citet{SHETH1995} who proposed to use a variable transformation in order to derive a generalized Wiener filter for the lognormal distribution. This approach, however, yielded a very complex form for the noise covariance matrix making applications to real data sets impractical. The first successful application of the lognormal Poissonian distribution for density inference was presented by \citet{SAUNDERS2000A}. Their method is based on the expansion of the density logarithm into spherical harmonics \citep[][]{SAUNDERS2000}. More accurate schemes based on maximum and mean posteriori principles were derived by \citep{ENSSLIN2008}. Recently, an implementation of the maximum a posteriori scheme was presented and thoroughly tested by \citep{KITAURA2009B}. They found that, assuming a linear bias, the lognormal Poissonian posterior permits recovery of the density field deep in the nonlinear regime up to values \(\delta \ge 1000\) of the density contrast. Finally, \citet{JASCHE2009B} developed the Hamiltonian density estimation and sampling scheme to map out the posterior probability distribution.

\subsection{HADES}
\label{HADES}
As already described above, Bayesian non-linear large scale structure inference requires to sample from non-Gaussian posterior distributions. In order to do so, we developed HADES \citep[see][ for more details]{JASCHE2009}. HADES explores the very high dimensional parameter space of the three dimensional density field via an Hamiltonian Monte Carlo (HMC) sampling scheme. Unlike conventional Metropolis Hastings algorithms, which move through the parameter space by a random walk, and therefore require prohibitive amounts of steps to explore high dimensional spaces, the HMC sampler suppresses random walk behavior by introducing a persistent motion of the Markov chain through the parameter space \citep[][]{DUANE1987,NEAL1993,NEAL1996}. In this fashion, the HMC sampler maintains a reasonable efficiency even for high dimensional problems \citep{HANSON2001}.
Since it is a fully Bayesian method, the scientific output is not a single estimate, but a sampled representation of the multidimensional lognormal Poissonian posterior distribution given in equation (\ref{eq:LOGNORMALPOISSONIAN_POSTERIOR}). Given this representation of the posterior any desired statistical summary, such as mean, mode or variances can easily be calculated. Further, any uncertainty can seamlessly be propagated to the finally estimated quantities, by simply applying the according estimation procedure to all Hamiltonian samples.
For a detailed description of the theory behind the large scale structure sampler and its numerical implementation see \citet{JASCHE2009}.

\begin{figure}
\centering{\includegraphics[width=0.6\textwidth,clip=true]{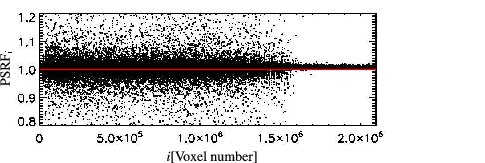}}
\caption{Results of the Gelman \& Rubin convergence diagnostic. The PSRF indicates convergence. As can be seen the Gelman \& Rubin test converges faster in regions with good data.}
\label{fig:gelmanrubin}
\end{figure}

\begin{figure*}
\centering{\includegraphics{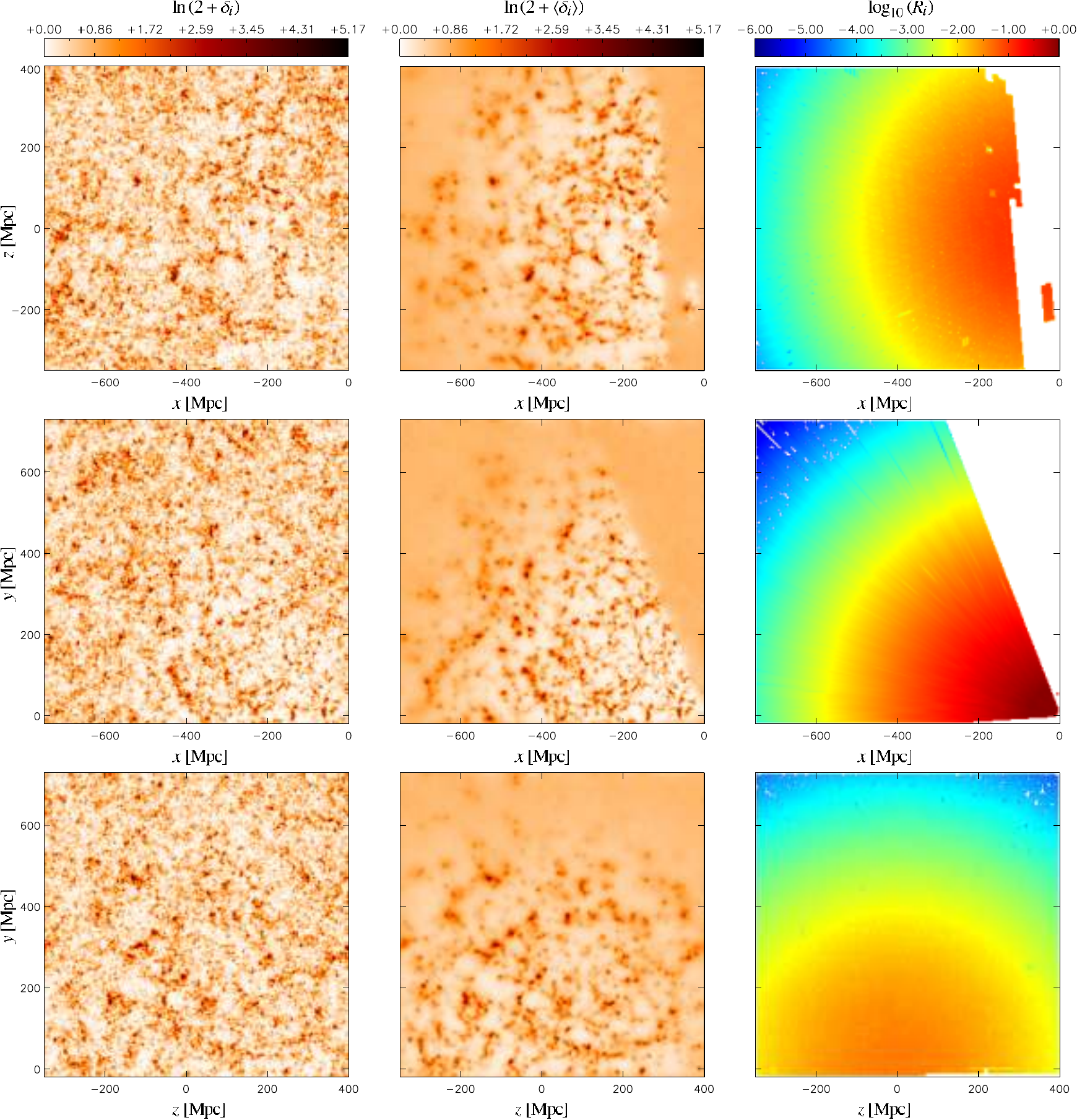}}
\caption{Three different slices from different sides through density fields. Left panels show slices through one of the \(40000\) density sample, middle panels depict the estimated ensemble mean and right panels demonstrate the according slices through the three dimensional response operator \(R_i\). It can be seen that the density sample (left panels) possesses equal power throughout the entire domain, even in the unobserved regions.}
\label{fig:densitysample_2000}
\end{figure*}

\subsection{Classification of the cosmic web}
\label{web_classification}
The results generated by the Hamiltonian sampler HADES will permit a variety of scientific analyses of the large scale structure in the observed Universe. An interesting example is to classify the cosmic web, in particular identifying different types of structures in the density field. Such an analysis, in example, is valuable for studying the environmental dependence of galaxy formation and evolution \citep[see e.g.][]{LEE2008,LEELI2008}. 
Since the structure classification is not always unique, we provide the full Bayesian posterior distribution of the structure type at a given position conditional on the observations. 

However, to do so we first need a means to identify different structure types from the density field. Numerous methods for structure analysis have been presented in literature \citep[see e.g.][]{LEMSON1999,COLBERG2005,NOVIKOV2006,HAHN2007,ARAGON2007,COLBERG2008A,FORERO2009}. In principle, all of these methods can be applied for the analysis of the Hamiltonian samples, however for the purpose of this paper we follow the dynamical cosmic web classification procedure as proposed by \citet{HAHN2007}. They propose to classify the large scale structure environment into four web types (voids, sheets, filaments and halos) based on a local-stability criterion for the orbits of test particles. The basic idea of this dynamical classification approach is that the eigenvalues of the deformation tensor characterize the geometrical properties of each point in space. The deformation tensor \(T_{ij}\) is given by the Hessian of the gravitational potential \(\Phi\):
\begin{equation}
\label{eq:DEFORM_TENSOR}
T_{ij}=\frac{\partial^2 \Phi}{\partial x_i\, \partial x_j} \, ,
\end{equation}
with \(\Phi\) being the rescaled gravitational potential given as \citep[see][]{FORERO2009}:
\begin{equation}
\label{eq:Poisson_eq}
\nabla^2 \Phi=\delta \, .
\end{equation}
It is important to note, that the deformation tensor, and the rescaled gravitational potential are both physical quantities, and hence their calculation requires the availability of a full physical density field in contrast to a smoothed mean reconstruction of the density field. As was already mentioned above, and will be clarified in section \ref{HMC_samples}, the Hamiltonian samples provide such required full physical density fields. The deformation tensor can therefore easily be calculated for each Hamiltonian sample from the Fourier space representation of equation (\ref{eq:DEFORM_TENSOR}).
Each spatial point can then be classified as a specific web type by considering the three eigenvalues, \(\lambda_1 \ge \lambda_2 \ge \lambda_3 \), of the deformation tensor. Namely, a void point corresponds to no positive eigenvalue, a sheet to one, a filament to two and a halo to three positive eigenvalues \citep[][]{FORERO2009}.
The interpretation of this rule is straight forward, as the sign of a given eigenvalue at a given position defines, whether the gravitational force at the direction of the principal direction of the corresponding eigenvector is contracting (positive eigenvalues) or expanding (negative eigenvalues). However, \citet{FORERO2009} found that rather than using a threshold value \(\lambda_{th}\) of zero different positive values can yield better web classifications. For this reason, in this work, we use the extended classification procedure as proposed by \citet{FORERO2009}.
The structures are then classified according to the rules given in table \ref{tb:structureclassification}. By applying this classification procedure to all Hamiltonian samples we are able to estimate the web type posterior \({\mathcal P}(\{{\rm T}_i(\vec{x}_k)\}|\{N_k^{g}\}, \lambda_{th} )\) of four different web types (\({\rm T}_1(\vec{x}_k)=\) void, \({\rm T}_2(\vec{x}_k)=\) sheet, \({\rm T}_3(\vec{x}_k)=\) filament, \({\rm T}_4(\vec{x}_k)=\) halo) conditional on the observations and the threshold criterion \(\lambda_{th}\). 

\begin{table}\centering
\begin{tabular}{| l | l |}
\hline
Structure type &  rule\\
\hline
Void & \(\lambda_1,\lambda_2,\lambda_3 < \lambda_{th}\)\\
Sheet & \(\lambda_1 > \lambda_{th}\) and \(\lambda_2,\lambda_3 < \lambda_{th}\)\\
Filament & \(\lambda_1,\lambda_2 > \lambda_{th}\) and \(\lambda_3 < \lambda_{th}\)\\
Halo & \(\lambda_1,\lambda_2,\lambda_3 > \lambda_{th}\)\\
\hline
\end{tabular}
\caption{Rules for the dynamic classification of web types.}
\label{tb:structureclassification}
\end{table}

\section{DATA}
\label{DATA}
In this section we describe the SDSS galaxy sample used for the analysis. Additionally, we discuss
the data preparation steps required to perform the three dimensional density inference procedure.

\subsection{The SDSS galaxy sample}

We use  data from {\tt Sample  dr72} of the New  York University Value
Added Catalogue (NYU-VAGC)
\footnote{http://sdss.physics.nyu.edu/vagc/}. This is an update of the
catalogue constructed by \citet{BLANTON2005} and is based on the
final data release \citep[DR7;][]{SDSS7}
of the Sloan Digital Sky Survey \citep[SDSS;][]{YORK2000}.
Starting from {\tt Sample dr72},
we   construct   a   magnitude-limited   sample   of   galaxies   with
spectroscopically  measured  redshifts  in  the  range  $0.001<z<0.4$,
$r$-band  Petrosian apparent magnitude  $r\leq 17.6$  after correction
for   Galactic    extinction,   and   $r$-band    absolute   magnitude
$-23<M_{^{0.1}r}<-17$. Here $M_{^{0.1}r}$  is corrected to its $z=0.1$
value using the $K$-correction code of 
\citet{BLANTON2003A} and \citet{BLANTON2007}
and the luminosity evolution model of 
\citet{BLANTON2003}.  The apparent
magnitude limit is chosen in order to get a sample that is uniform and
complete  over  the entire  area  of  the  survey.  We  also  restrict
ourselves  to galaxies  located in  the  main contiguous  area of  the
survey in the northern Galactic cap, excluding the three survey strips
in the  southern cap (about 10 per  cent of the full  survey area). In
addition, we consider only galaxies which are inside a comoving cube of
side length \(750\) Mpc. These restrictions result in a
final sample of 463,230 galaxies.

\begin{figure}
\centering{\includegraphics{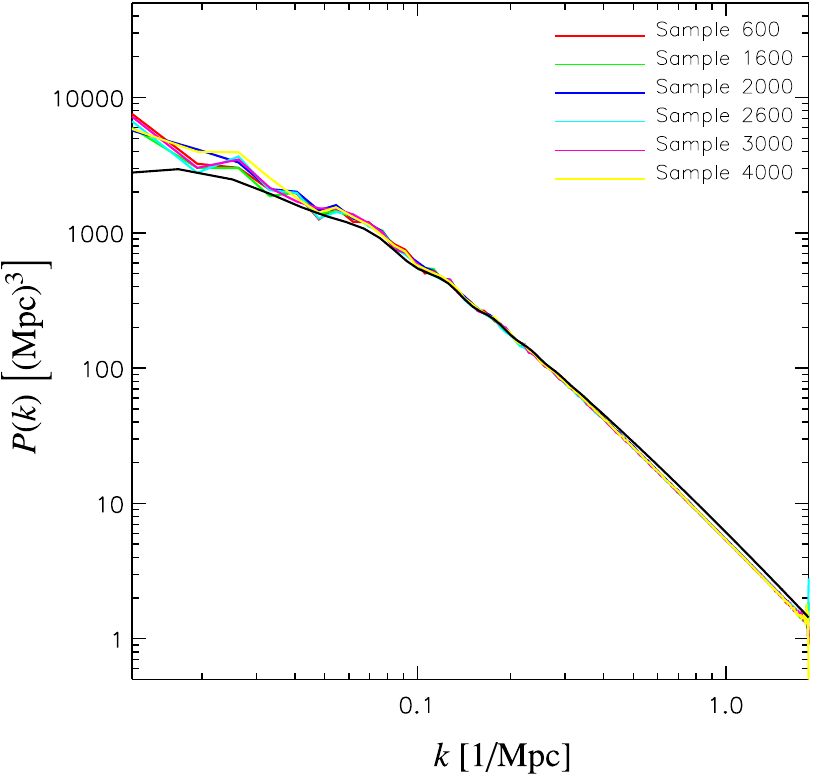}}
\caption{Power-spectra of some Hamiltonian samples. The black curve corresponds to a linear \(\Lambda\)CDM power-spectrum.}
\label{fig:power-spectra}
\end{figure}
The NYU-VAGC  also provides the  necessary information to  correct for
incompleteness in our spectroscopic sample. This includes two parts: a
mask which shows  which areas of the sky have  been targeted and which
have not, and a radial  selection function which gives the fraction of
galaxies  in the absolute  magnitude range  being considered  that are
within the apparent magnitude range of the sample at a given redshift.
The mask defines the effective area of the survey on the sky, which is
6437 deg$^2$ for  the sample we use here. This  survey area is divided
into a  large number of  smaller subareas, called {\it  polygons}, for
each of which the NYU-VAGC lists a spectroscopic completeness, defined
as the  fraction of photometrically identified target  galaxies in the
polygon for  which usable spectra  were obtained. Over our  sample the
average completeness is 0.92.

\subsection{Completeness and selection function}
Three dimensional density field inference requires the definition of the linear observation response operator \(R_k\), as given in section \ref{LOGNORMALPOISSONIANPOSTERIOR}. This response operator describes to what percentage each volume element of the three dimensional domain has been observed. It is hence a projection of the product of radial and angular selection function into the three dimensional voxelized space. In particular, we have to solve the convolution integral:
\begin{equation}
\label{eq:window_function}
R_k= R(\vec{x}_k)=\int \mathrm{d}\vec{y}\,W\left(\vec{x}_k-\vec{y}\right) \, f\left(r(\vec{y})\right)\,M\left(\alpha(\vec{y}),\delta(\vec{y})\right) \, ,
\end{equation}
where \(W(\vec{x})\) is the voxel kernel, \(f(r)\) is the radial selection function, with \(r\) being the distance from the observer and \(M(\alpha,\delta)\) is the angular selection function, where \(\alpha\) and \(\delta\) are right ascension and declination respectively.
We evaluate this integral numerically for the nearest grid point kernel by following different line of sights and calculating the contribution of the product of angular and radial selection function to each voxel.

As mentioned above, in this work we used the two dimensional sky mask and the radial selection function provided by the NYU-VAGC.
\begin{figure*}
\centering{\includegraphics{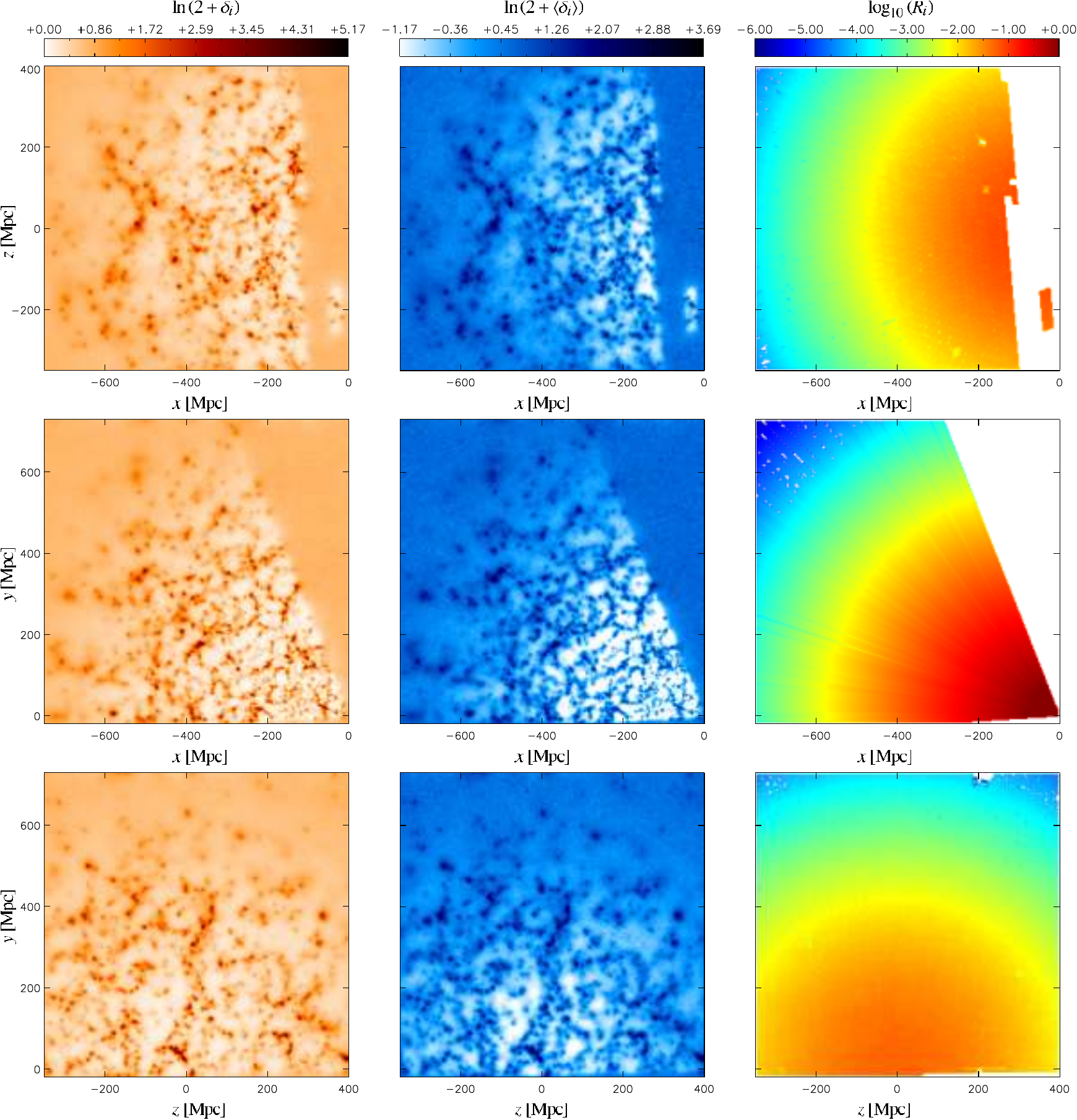}}
\caption{Three different slices from different sides through ensemble mean density (left panels), ensemble variance (middle panels) and the three dimensional response operator \(R_i\) (right panels). Especially the variance plots demonstrate, that the method accounted for the full Poisonian noise structure introduced by the galaxy sample. One can also see the correlation between high density regions and high variance regions, as expected for Poissonian noise.}
\label{fig:mean_variance_mask}
\end{figure*}

\subsection{Creating the three dimensional data cube}
\label{Coordinates}
The large scale structure sampler operates on a three dimensional equidistant grid. In particular, in this work we set up a cubic grid with side length \(750 \rm{ Mpc}\) and \(256^3\) voxels. This amounts to a resolution of \(\sim 3 \rm{ Mpc}\) voxel side length.
Since our algorithm relies on the correlation function in comoving space, all calculations are performed with comoving length units rather than with redshift distances. For this reason, we transform all galaxy redshifts \(z\) to comoving distances via the relation:
\begin{equation}
\label{COM_COORDS}
r=\int_0^{z_i} dz \frac{1}{c\,H(z)} \,
\end{equation}
where \(z_i\) is the estimated galaxy redshift, \(c\) is the speed of light and \(H(z)\) is the Hubble parameter given as:
\begin{equation}
\label{HUBBLE_FUNCTION}
H(z)=H_0\sqrt{\Omega_m\,(1+z)^3 + \Omega_c\,(1+z)^2+\Omega_{\Lambda}}\, .
\end{equation}
Further, we choose a concordance \(\Lambda\)CDM model with a set of cosmological parameters \((\Omega_m=0.24\, ,\, \Omega_c=0.00 \, , \, \Omega_{\Lambda}=0.76\, , \,  h=0.73  \, , \,  H_0=h\, 100\,{\rm{km/s/ Mpc}}  )\) \citep[][]{SPERGEL2007}.
With these definitions we can calculate the three dimensional cartesian coordinates for each galaxy as:
\begin{eqnarray}
\label{CARTESIAN_COORDS}
x &=& r\,\cos(\delta)\,\cos(\alpha) \nonumber \\
y &=& r\,\cos(\delta)\,\sin(\alpha) \nonumber \\
z &=& r\,\sin(\delta) \, 
\end{eqnarray}
where \(\alpha\) and \(\delta\) are the right ascension and declination respectively. We then sort the galaxy distribution into the three dimensional equidistant grid via a nearest grid point procedure \citep[see e.g.][]{HOCKNEYEASTWOOD1988}. An estimate for the expected number of galaxies \(\bar{N}\) can then be calculated as:
\begin{equation}
\label{HUBBLE_FUNCTION}
\bar{N}=\frac{\sum_k N_k^{g} }{\sum_l R_l}\, 
\end{equation}
\citep[see e.g.][for details]{KITAURA2009,JASCHE2009}.

\subsection{Physical model}
\label{physical_model}
Observations of the galaxy redshifts do not permit direct inference of the underlying matter distribution. Various physical effects such as galaxy biasing and redshift space distortions must be taken into account for proper analyses. This is of particular relevance for the choice of power-spectrum required for the sampling procedure (see equation (\ref{eq:LogNormal_prior})).
However, according to the discussion in \citet{ERDOGDU2004} and \citet{KITAURA2009} these effects can be greately accounted for in a separate postprocessing step, once the continuous expected galaxy density field in redshift space has been obtained.
For this reason, here we seek to recover the density field in redshift space permitting us to test various bias models and redshift space distortions correction methods in a subsequent step.
 
In particular, the relation between the true underlying dark matter density field and the expected continuous galaxy density contrast is generally very complicated and involves non-local and non-linear bias operators. Several non-local bias models have been presented, which mostly aim at correcting the large scale power in power-spectrum estimation procedures \citep[see e.g.][]{sdss-tegmark,SELJAK2000,PEACOCKSMITH2000,HAMANN2008}. As described in section \ref{Methodology} and \ref{HADES} the Hamiltonian sampler is able to account for such bias models.
Note however, that a specific bias model also fixes the model for the underlying dark matter distribution.
Therefore, here, we prefer to follow the approach of previous works of setting
the bias operator to a constant linear factor equal to unity \citep[][]{ERDOGDU2004,KITAURA2009}. In this fashion, one obtains the expected continuous galaxy density contrast. As discussed in \citet{KITAURA2009}, the according underlying dark matter distribution can then be simply obtained by deconvolving the results with a specific scale dependent bias model, permitting us to explore various different bias models.

In a similar manner, one can treat redshift-space distortion effects. These are mainly due to the peculiar velocities of galaxies, which introduces Doppler effects in the redshift measurement \citep[see e.g.][]{1987MNRAS.227....1K,1994MNRAS.267.1020P,hamilton-1998,DAVISPEEBLES1983}. This effect leads to a radial smearing of the observed density field in redshift-space and yields elongated structures along the line of sight, the so called \textit{finger-of-God} effect.
% The observed redshift space density field therefore is the convolution of its real-space counterpart with the radial peculiar velocity distribution of galaxies.

Additionally, there exists a cosmological redshift-space effect which is sensitive to the global geometry of the Universe. In particular, the comoving separation of a pair of galaxies at \(z \gg 0.1\) is not determined only by their observable angular and redshift separations without specifying the geometry, or equivalently the matter content of the Universe \citep[][]{MAGIRA2000}. This effect yields anisotropies in the matter distribution especially at \(z\ge1\) \citep[see e.g.][]{AP-79,MATSUBARA1996,1996MNRAS.282..877B,Popowski1998}.
However, for the volume considered in this work (\(z\le 0.27\)), the dominant
redshift-space distortions are due to non-linear peculiar motions
of galaxies in large overdensities.
This effect has pronounced consequences for the power-spectrum in redshift-space, since it suppresses power on small scales. As demonstrated in \citet{ERDOGDU2004}, the redshift-space power-spectrum of a fully evolved non-linear matter distribution is very similar to a linear power-spectrum at the scales relevant for this work (\(k \le 2 \) \(\rm{h/Mpc}\)). Here, they used the non-linear power-spectrum fitting formula provided by \citet{SMITH2003}. However, the exact galaxy power-spectrum in redshift-space is not known. The work of \citet{TEGMARK2006} indicates that the recovered power-spectrum of the SDSS main sample is close to a linear power-spectrum, which may be due to the fact that this galaxy sample is not strongly clustered. In this case, the redshift-space power-spectrum would be even closer to a linear power-spectrum. In any case assuming a linear power-spectrum will still permit physically accurate matter field inference in redshift-space \citep{ERDOGDU2004}. For this reason, in the absence of more precise information on the galaxy power-spectrum in redshift-space, here we will assume a linear power-spectrum, calculated according to the prescription provided by \citet{1998ApJ...496..605E} and \citet{1999ApJ...511....5E}. One should also bear in mind that the data itself will govern the inference process. For this reason, power-spectra measured from the Hamiltonian samples will only be partially defined by the a priori power-spectrum guess but mostly by the data. However, we defer a more careful treatment of all physical effects including a joint inference of density field and power-spectrum to a future work.

It is clear, that precise
correction of these redshift-space effects requires knowledge
about the peculiar velocities of all observed galaxies, which is
usually not provided by galaxy redshift surveys.
Therefore, precise correction of redshift-space distortions is very complicated and subject to ongoing research.
In the linear regime, the theory behind the observed redshift-space distortions is well developed \citep[][]{1987MNRAS.227....1K,hamilton-1998}. However, in quasi-linear and non-linear regimes, we instead have to resort to making approximations or using fitting formulae based on numerical simulations \citep[][]{PERCIVAL2009}.
Literature provides numerous approaches to alleviate these redshift-space distortions particularly in power-spectrum estimation. Most of these approaches aim at restoring the correct power by deconvolution with an redshift-space convolution kernel which takes into account the random pair velocities of galaxies in collapsed objects \citep[see e.g][]{1994MNRAS.267.1020P,1996MNRAS.282..877B,JINGBOERNER1998,hamilton-1998,KANG2002,JINGBOERNER2004,ERDOGDU2004,SCOCCIMARRO2004,CABRE2009,PERCIVAL2009}. Such techniques have been adopted to correct Wiener density reconstructions by applying a redshift distortion operator to the final result, in order to restore the correct power \citep[][]{ERDOGDU2004,KITAURA2009}. However, it must be noted that this method does not account for the correction of phase information, and therefore only corrects the two-point statistics of the recovered density field.

Three dimensional density inference hence requires redshift-space
distortions corrections which also account for phase information and would
be dependent on the density or gravitational potential. In the linear
regime it is possible to apply an inverse redshift-space operator
which transforms the redshift-space density to its real-space
counterpart \citep[][]{TAYLOR1999,DMELLOW2000}. However, it does not account for the strongly non-linear regime which mostly generates the \textit{finger-of-God} effect. For this reason \citet{sdss-tegmark} proposed a \textit{finger-of-God} compression method. Here they use a standard friends-of-friends algorithm to identify a cluster by taking into account different density thresholds, which set the linking length. They then measure the dispersion of galaxies about the cluster center along the line of sight and in transverse direction. If the radial dispersion exceeds the transverse dispersion, the cluster is compressed radially until the radial dispersion equals the transverse dispersion \citep[][]{sdss-tegmark}. However, it is not clear to what degree such a method would falsely isotropize filaments or under dense objects along the line of sight to spherical clusters. Such a method of isotropizing the density field, however, can also be applied in a post processing step, by noting that a density threshold refers to a linking length in the friends-of-friends algorithm. 

Nevertheless, the above correction methods mask the fact that redshift-space distortions introduce statistical uncertainties. Thus unique recovery of the real-space density field is generally not possible. A full characterization of the joint uncertainties of the real-space density hence would require to carefully take into account the uncertainties introduced by redshift-space distortions or the lack of knowledge on peculiar velocities. This can be achieved by introducing a density dependent peculiar velocity sampling scheme to our method, as proposed by \citet{Kitaura}. However, we defer sampling of the peculiar velocities to a future
work.

\begin{figure}
\centering{\includegraphics{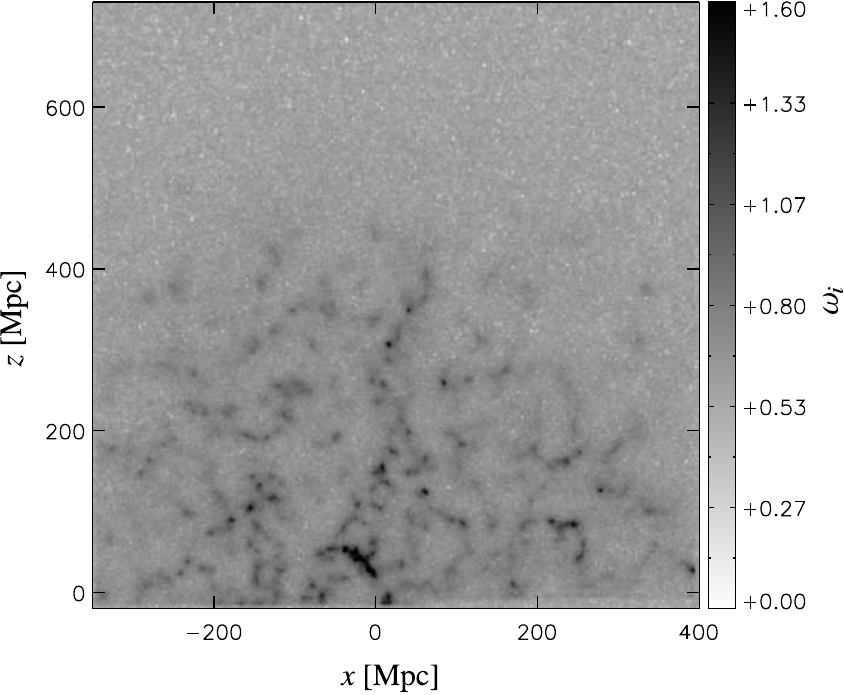}}
\caption{The plot shows the relative density to variance ratio \(\omega_i\). In comparison to the lower panels of figure \ref{fig:mean_variance_mask} it indicates a high signal-to-noise ratio in regions of high density as expected for Poissonian noise.}
\label{fig:rel_dens_var}
\end{figure}

\section{Results}
\label{RESULTS}
In this section we describe the results obtained from the large scale structure inference procedure. % The result is not a single estimate, but a high dimensional point cloud consisting of individual Hamiltonian samples, which represent the full non Gaussian lognormal Poissonian posterior. The advantages of possessing a representation of the large scale structure posterior are manifold. It enables us to calculate any statistical summary, such as mean, mode and variances. Further, we are able to estimate the full non-Gaussian uncertainties, and non-linear, non-Gaussian error propagation to any finally estimated quantity can be simply achieved by applying the estimation procedure to all Hamiltonian samples. It is also possible to provide full probability distributions for derived quantities, such as classified objects as presented later in this work.

\subsection{Convergence test}
\label{CONVTEST}
HADES is a Markov Chain Monte Carlo sampler and hence we have to test, if the individual Hamiltonian samples really represent the lognormal Poissonian posterior. Convergence diagnostic of Markov chains is subject of many discussions in literature \citep[see e.g.][]{HEIDELBERGER1981,GELMAN1992,GEWEKE1992,RAFTERY1995,COWLES1996,HANSON2001,DUNKLEY2005}. However, here we apply the widely used Gelman \& Rubin diagnostic, which is based on multiple simulated chains by comparing the variances within each chain and the variance between chains \citep{GELMAN1992}. In particular, we calculate the potential scale reduction factor (PSRF) \citep[see][]{JASCHE2009B}. A large PSRF indicates that the inter chain variance is substantially greater than the intra chain variance and longer chains are required. Once the PSRF approaches unity, one can conclude that each chain has reached the target distribution. We calculated the PSRF for each voxel in our calculation domain. The result for this test is presented in figure \ref{fig:gelmanrubin}. It indicates convergence of the Markov chain. However, it can be seen that some regions of the domain converge faster than others. This is due to the fact, that not all regions of the cubical volume have been observed equally. Regions which contain good observations converge faster, since there the probability distribution is narrower, while poorly or non observed regions converge slower, since the space of possible solutions is larger.
Also note, that the Gelman \& Rubin diagnostic is generally a conservative test, and other tests might indicate convergence much earlier. However, this test clearly demonstrates that the quality and amount of observational data can have a strong impact on the convergence behavior of the chain.

\subsection{Hamiltonian samples}
\label{HMC_samples}
Since the Markov chain converges we can conclude, that the individual samples are really samples from the large scale structure posterior.
At this point it is important to insist that the Hamiltonian samples are not the result of a filtering procedure. A filter generally suppresses the signal in low-signal to noise regions, and therefore produces biased estimates for the physical density field. This is not the case for the individual Hamiltonian samples. Since they are random realizations of the lognormal Poissonian posterior, they are unbiased density fields in the sense that they possess correct physical power throughout the entire cubical volume. As an example we present slices through an arbitrary density sample in figure \ref{fig:densitysample_2000}. Already visually, one has the impression, that the density field has equal power throughout the entire domain, even in the unobserved regions. This is because the Hamiltonian sampler non-linearly augments the poorly or not observed regions with statistically correct information. Each density sample therefore is a proper physical density field, from which physical quantities can be derived. To demonstrate this, we measure the power-spectra of some of these Hamiltonian samples. The result is presented in figure \ref{fig:power-spectra}. As can be seen, the power-spectra of the individual samples, are very close to the assumed linear \(\Lambda\)CDM power-spectrum. The deviations at large scales and small scales are due to the impact of the data. At small scales the deviation can be explained by redshift space distortions, while at the largest scales cosmic variance is dominant. There is clearly no sign of artificial power loss due to the survey geometry.
Since the individual samples are valid density field realizations, it is easy to derive meaningful physical quantities, such as the gravitational potential, cosmic flows or the tidal shear tensor as demonstrated in the remainder of this paper.

\begin{figure}
\centering{\includegraphics{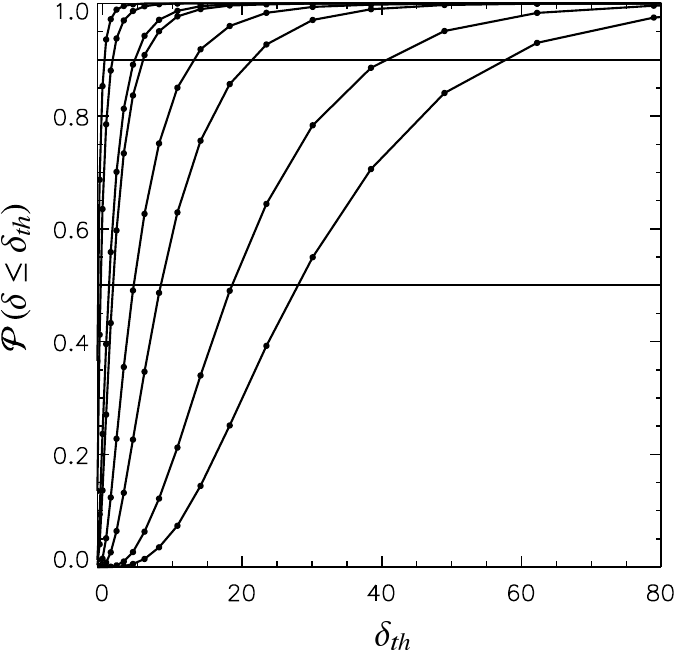}}
\caption{Commulative probability distributions of the density at randomly chosen points in the volume. The cumulative probability distributions have been evaluated for 20 threshold values \(\delta_{th}\). The two horizontal lines indicate the \({\mathcal P}\left(\delta \le \delta_{th}\right)=0.5\) and \(0.9\) thresholds respectively.}
\label{fig:comprob}
\end{figure}

\subsection{Ensemble mean and variance}
\label{MEANANDVARIANCE}
%As described in the previous sections, the scientific output of our method is a sampled representation of the lognormal Poissonian posterior, which allows us to calculate any desired statistical summary.
%In particular,
Here we want to present the ensemble mean and variance for the set of \(40 000\) Hamiltonian samples, each consisting of \(256^3\) voxels.
%We estimate the ensemble mean and variance from a set of \(40 000\) Hamiltonian samples, each consisting of \(256^3\) voxels.
For comparison with a single density sample the middle panels of figure \ref{fig:densitysample_2000} show the according slices through the ensemble mean density field, which exhibits many interesting features. First of all, it renders remarkably well the filamentary structure of our cosmic neighborhood. Many clusters, filaments and voids can clearly be seen by visual inspection.
In the unobserved regions the ensemble mean density amplitudes drop to the cosmic mean for the density contrast \(\delta=0\), just as required by construction. Structures close to the observer, at cartesian coordinates \( (0,0,0)\), are more clearly visible than structures at larger distances. Especially, filaments and voids are less prominent at larger distances. This is due to the observational response operator \(R_i\), which due to the radial selection function drops to very low values at large distances. Therefore, once a galaxy is detected far away from the observer, it must reside inside a large overdensity, and hence inside a cluster. This expectation is clearly represented by the ensemble mean density field.
Another interesting point to remark is, that the borders to the unobserved regions are not very sharp. Some of the observed information is non-linearly propagated into the unobserved regions, since our method takes into account the correlation structure of the underlying signal. It can therefore be seen, that some clusters and voids are interpolated further out into the unobserved regions. In comparison to the Wiener filter as previously applied to SDSS data by \citet{KITAURA2009}, it seems that the Hamiltonian sampler is more conservative and 
less optimistic for the extrapolation of information into the unobserved region. This may be due to the fact, that here we take into account the full Poissonian noise statistics rather than restricting the noise to a Gaussian approximation.
Beside the ensemble mean, here we also calculate the ensemble variance per voxel, which is the diagonal of the full ensemble covariance matrix. Some slices through the ensemble mean, ensemble variance and the according slices through the observational response operator are presented in figure \ref{fig:mean_variance_mask}. Here the middle panels correspond to ensemble variance. At first glance, one can nicely see the Poissonian nature of the galaxy shot noise. High density peaks in the ensemble mean map correspond to high variance regions in the ensemble variance map, as expected for Poissonian noise. One can clearly see that the Hamiltonian sampler took into account the full three dimensional noise structure of the galaxy distribution. Additionally, with larger distance to the observer, the average variance increases, as is expected due to the radial selection function.
It is also interesting to remark, that some voids have been detected with quite low variance, hence with high confidence.
Note, however, although here we only plotted the diagonal of the density covariance matrix, the full non-diagonal covariance structure is completely encoded in the set of Hamiltonian samples, and can be taken into account for future analysis.
Also note, that the variance slices show high variances in regions where many galaxies have been observed. This is a key feature of the Poisson statistics, because the standard deviation is equal to the square-root of the number of individual galaxies. That is, if there are \(N\) galaxies in each voxel, the mean is equal to \(N\) and the standard deviation is equal to \(\sqrt{N}\). This makes the signal-to-noise ratio equal to \(\sqrt{N}\) for such an homogeneous case.
To emphasize the fact, that regions which show high variances have also high signal-to-noise ratios, we calculate the density to variance ratio:
\begin{equation}
\label{eq:dens_var_ratio}
\omega_i = \frac{\left(1+ \langle \delta_i\rangle\right)}{\sqrt{\langle \delta_i^2 \rangle - \langle \delta_i \rangle^2 }} \, .
\end{equation}
The result of this calculation is presented in figure \ref{fig:rel_dens_var} for the case of the lower slices of figure \ref{fig:mean_variance_mask}. It clearly indicates high signal-to-noise ratios in high density regions.
\begin{figure*}
\centering{\includegraphics{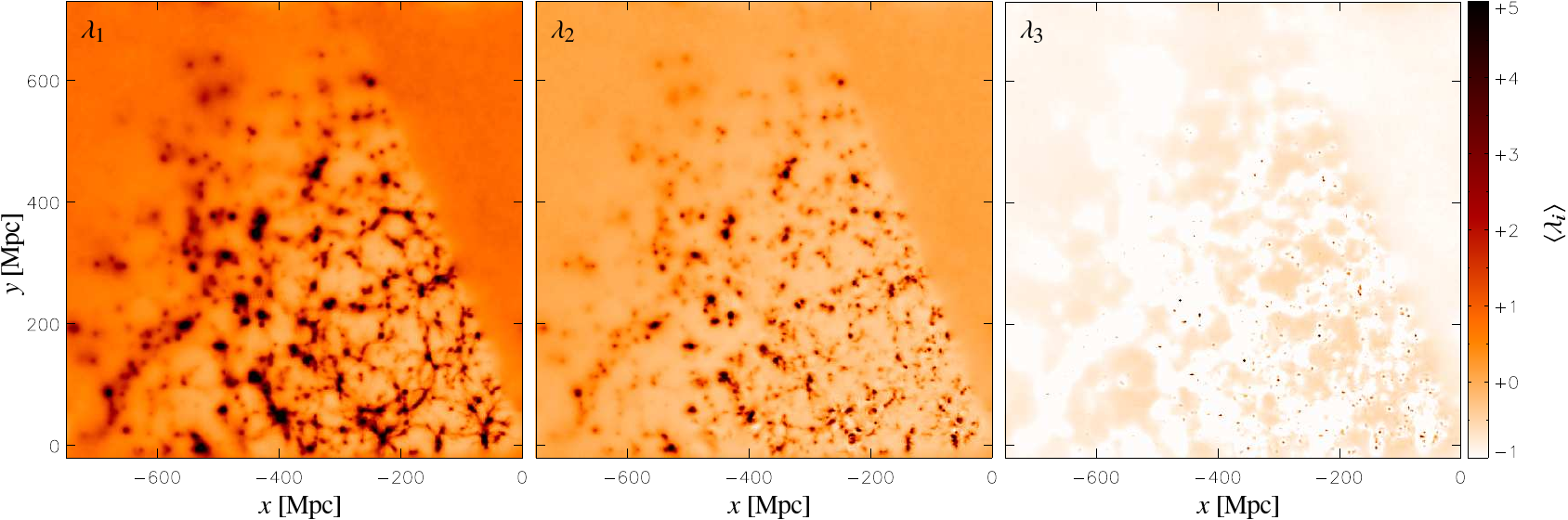}}
\caption{Ensemble mean of the eigenvalues of the deformation tensor.}
\label{fig:picture9}
\end{figure*}
In addition, we also estimate the cumulative probabilities \({\mathcal P}\left(\delta_i \le \delta_{th}\right)\), at twenty different density threshold values \(\delta_{th}\), for the density found at each voxel. This cumulative probabilities are estimated from the Hamiltonian samples by: 
\begin{equation}
\label{eq:cumulative_prob}
{\mathcal P}\left(\delta_i \le \delta_{th}\right)=\frac{\sum_{n=1}^{N_{\mathrm{samp}}} \Theta(\delta_{th}-\delta_i) }{N_{\mathrm{samp}}} \, ,
\end{equation}
where \(n\) labels the individual Hamiltonian samples, \(N_{\mathrm{samp}}\) is the total number of samples and \(\Theta(x)\) is the Heaviside function.
These cumulative probabilities allow for example to estimate the median density at each voxel, and can be usefull, when analyzing galaxies in their cosmic environment as will be done in a following project. Some such cumulative probability distributions, chosen randomly, are shown in figure \ref{fig:comprob}. As can be seen, the recovered density amplitudes extend over a large range, from small linear to very high non-linear values.

\section{Web classification}
\label{web_classification_data}
Already in the introduction we mentioned that the results presented in section \ref{RESULTS} are to be used for analyzing galaxy properties in the large-scale environment in a future work. Such analyses also require the classification of the large scale density field into different web type objects. Therefore, in order to characterize the environment of our SDSS galaxy population, here we apply the dynamic web classification procedure, as described in section \ref{web_classification}, to the set of Hamiltonian samples. A similar analysis has been previously carried out by \citet{LEEANDERDOGDU2007} and \citet{LEE2008} based on a Wiener mean density reconstructions of the 2MASS Redshift survey to study alignments of galaxy spins with the tidal field and the variation of galaxy morphological type with environmental shear.

Here we will follow a similar procedure to classify each individual voxel of a given Hamiltonian sample as one of the four web types \({\rm T}_i\), with the different types being \({\rm T}_1=\) void, \({\rm T}_2=\) sheet, \({\rm T}_3=\) filament, \({\rm T}_4=\) halo. To do so, we perform the following three steps for an individual Hamiltonian sample:
\begin{enumerate}
     \item {Solve equation (\ref{eq:DEFORM_TENSOR}) for the deformation tensor \(T_{ij}\) by means of Fast Fourier Transform techniques}
     \item {Solve the cubic characteristic equation for the three eigenvalues of the deformation tensor at each spatial position}
     \item {Apply the rules given in table \ref{tb:structureclassification} to classify the web type at each spatial position for a given threshold value \(\lambda_{th}\).}
\end{enumerate}
The result of this procedure for the \(n\)th sample is then a unit four vector \( \vec{{\rm T}}^n(\vec{x}_k)\) at each voxel position \(\vec{x}_k\). All of the entries of this four vector are zero except for one, which indicates the web type.
Applying the above method to all Hamiltonian samples will yield a set of classification four vectors, which encodes the information and uncertainty of the observations. Additionally, as an intermediate result, we obtain the set of the three eigenvalues for each individual Hamiltonian sample. Slices through their ensemble mean estimates are presented in figure \ref{fig:picture9}. 

However, rather than summarizing the results in terms of mean and variance here we want to estimate the full cosmic web posterior. This is achieved by counting the relative frequencies for each web type at each individual spatial coordinate within the set of Hamiltonian samples. With these definitions we yield the cosmic web posterior for each web type as:
\begin{equation}
\label{eq:web_type_posterior}
{\mathcal P}\left({\rm T}_i(\vec{x}_k)|\{N_k^{g}\}, \lambda_{th} \right)=\frac{\sum_{n=1}^{N_{\mathrm{samp}}} \sum_{j=1}^4 \delta^K_{{\rm T}_i(\vec{x}_k)\,{\rm T}^n_j(\vec{x}_k)} }{N_{\mathrm{samp}}} \, ,
\end{equation}
where \(n\) labels the individual Hamiltonian samples, \(N_{\mathrm{samp}}\) is the total number of samples and \(\delta^K_{ij}\) is the Kronecker delta.
% This result is especially appealing from a Bayesian point of view, since it emphasizes the fact, that the result of a Bayesian analysis is not a single estimate but rather an entire probability distribution function.
The cosmic web posterior incorporates all observational information and uncertainties, and enables us to determine how well different structures can be classified with respect to observational uncertainties.

\begin{figure*}
\centering{\includegraphics[width=1.0\textwidth,clip=true]{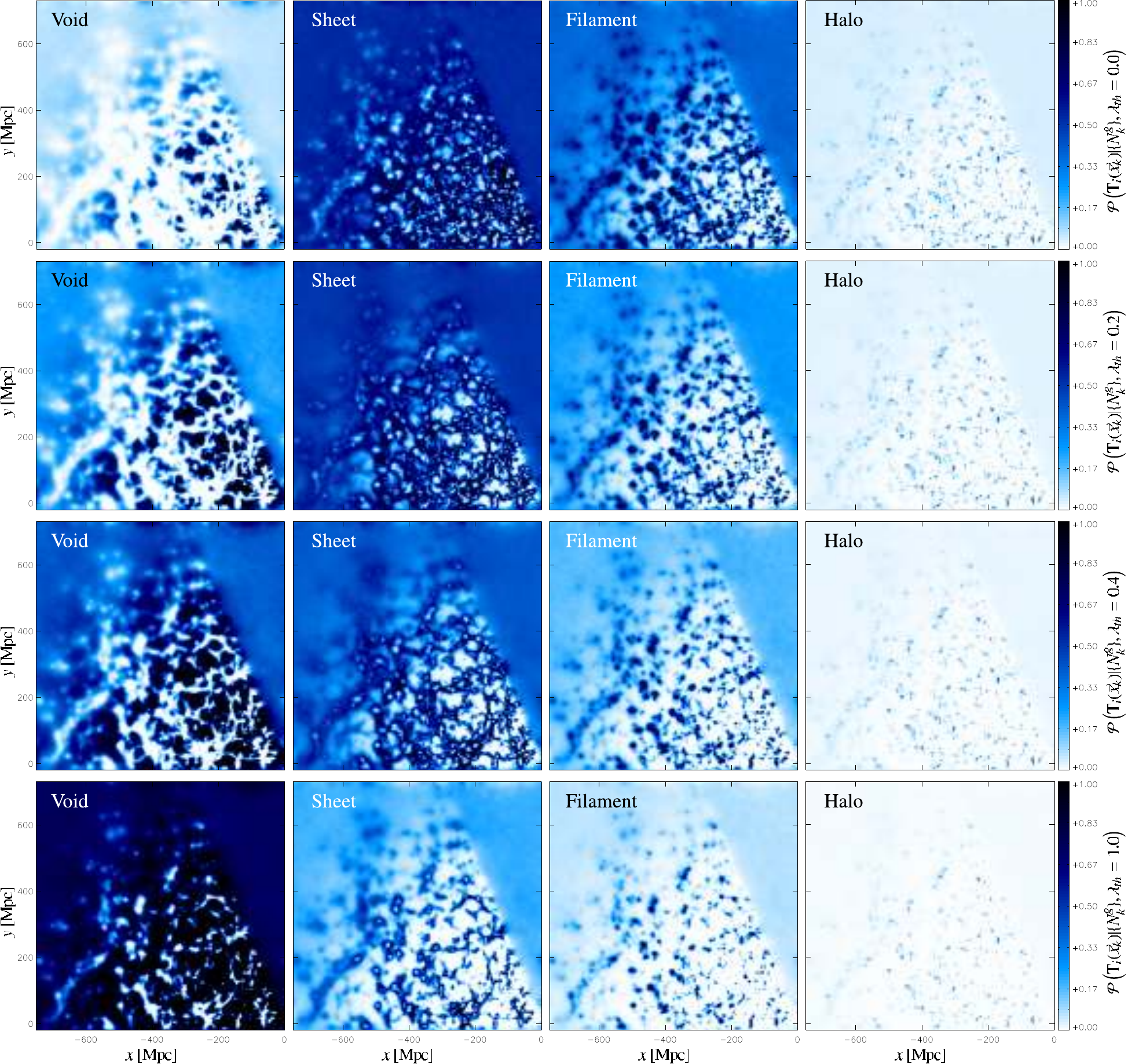}}
\caption{Slices through the cosmic web posterior for the threshold values \(\lambda_{th}=0.0,0.2,0.4,1.0\) (from top to bottom) for the four different web types. It is interesting to note, that sliced sheets look filamentary, while filaments piercing the slice appear as dots.}
\label{fig:webposterior_a}
\end{figure*}

\begin{figure*}
\centering{\includegraphics[width=0.8\textwidth,clip=true]{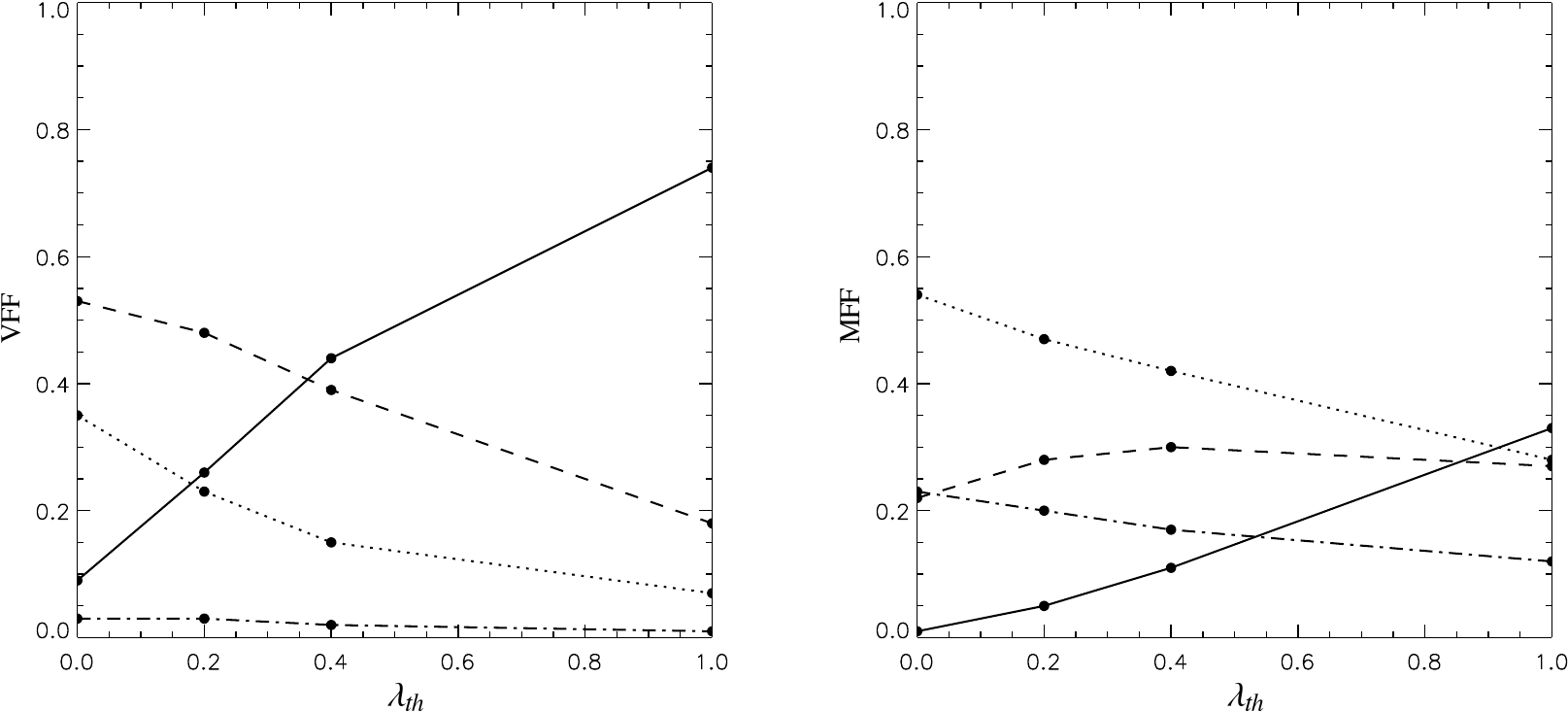}}
\caption{Volume and Mass filling factors as a function of \(\lambda_{th}\). Continuous lines denote voids, dashed lines sheets, dotted lines filaments and dot-dashed lines halos. Especially the void VFF and MFF respond strongly to a change in \(\lambda_{th}\) making them a sensitive measure of the cosmic web \citep[][]{FORERO2009}. }
\label{fig:VFF_MFF}
\end{figure*}

We evaluate the cosmic web posterior for four different values of \(\lambda_{th}\), with \(\lambda_{th}=0.0, 0.2, 0.4, 1.0\). Slices through the cosmic web posteriors for the four different cases are presented in figure \ref{fig:webposterior_a}. It can be clearly seen, that the properties of the survey geometry are represented by the four posterior distributions. While the web classification in the observed regions clearly follows the structure of the underlying density field, it obviously can not provide a clear classification of unobserved regions. Also with distance to the observer, the web classification becomes more and more uncertain. In this fashion, the cosmic web posterior renders the uncertainties introduced by the radial selection function and the resulting higher shot noise contribution at larger distances. 
The impact of the \(\lambda_{th}\) threshold can be observed when comparing the four cosmic web posteriors. In the case of \(\lambda_{th}=0.0\) the cosmic web consists of many small isolated voids, which occupy only a small fraction of the total area of the slice.  With increasing threshold \(\lambda_{th}\), voids become bigger and more connected until they completely dominate the cosmic web for the case \(\lambda_{th}=1.0\).
The opposite behavior can be observed in case of the halo posteriors, as the number of clearly detected halos declines with increasing threshold \(\lambda_{th}\).
Following \citet{FORERO2009}, we also calculate the volume occupied by
each web type (the volume filling fraction - VFF) and the fraction of
mass contained in such a volume (mass filling fraction - MFF). The
results are presented in figure \ref{fig:VFF_MFF}, and show the same
behavior as described in \citet{FORERO2009}. Figure \ref{fig:VFF_MFF}
supports the visual impression, gained by inspection of figure
\ref{fig:webposterior_a}, that especially the VFF and MFF for voids
strongly depend on the threshold value \(\lambda_{th}\). This shows
that voids can serve as a sensitive monitor and indicator of the
cosmic web \citep[][]{FORERO2009}. Unfortunately, \citet{FORERO2009}
do not provide an explicit gauging of the \(\lambda_{th}\) values from
simulations. Such a gauging and hence a clear definition of the
different cosmic web types would be very valuable for these types of analysis.

Having now a representation of the web type posterior we can for example calculate the odds \(O_i(\vec{x}_k)\) ratio given as:
\begin{equation}
\label{eq:posterior_odds}
O_i(\vec{x}_k)=\frac{{\mathcal P}\left({\rm T}_i(\vec{x}_k)|\{N_k^{g}\}, \lambda_{th} \right)}{1-{\mathcal P}\left({\rm T}_i(\vec{x}_k)|\{N_k^{g}\}, \lambda_{th} \right)} \frac{1-{\mathcal P}\left({\rm T}_i(\vec{x}_k) \right)}{{\mathcal P}\left({\rm T}_i(\vec{x}_k) \right)} \, ,
\end{equation}
which tells us how much a specific web type is favored over all others. Here, the \({\mathcal P}\left({\rm T}_i(\vec{x}_k) \right)\) can be obtained by averaging over all voxels in the volume. In example, this permits us to build a simple structure type map \(m(\vec{x}_k)\) which can be used for visual analyses as presented in the next section. Such a map can be defined as:
\begin{eqnarray}
\label{eqn:web_type_map}
m(\vec{x}_k) = \left \{ \begin{array}{ll}
  {\rm T}_i(\vec{x}_k) & \quad \mbox{for $O_i(\vec{x}_k) \ge O_{th}$}\\
  undecided & \quad \mbox{else}\\ \end{array} \right. \, ,
\end{eqnarray}
where \(O_{th}\) is an odds threshold usually chosen larger than unity.

\section{Galaxy properties versus LSS}
\label{GALAXY_PROPERTIES}
In this  section we present a preliminary,  but intuitive examination
of the  correlations between  the large-scale environment  of galaxies
and  their physical properties.   Here we  consider two  properties of
galaxies: stellar mass $M_\star$ and  $g-r$ color, and study how these
are  correlated  with  the  overdensity $\delta$  of  the  large-scale
environment  and its  type, which  is  one of  the four  web types  as classified as
halo,  filament, sheet, and void.  We  will come back
to  this  topic in  a  separate  paper  by considering  more  physical
properties of  galaxies and  performing more careful  and quantitative
analyses.

\begin{figure*}
\centering{\includegraphics[width=1.05\textwidth,clip=true]{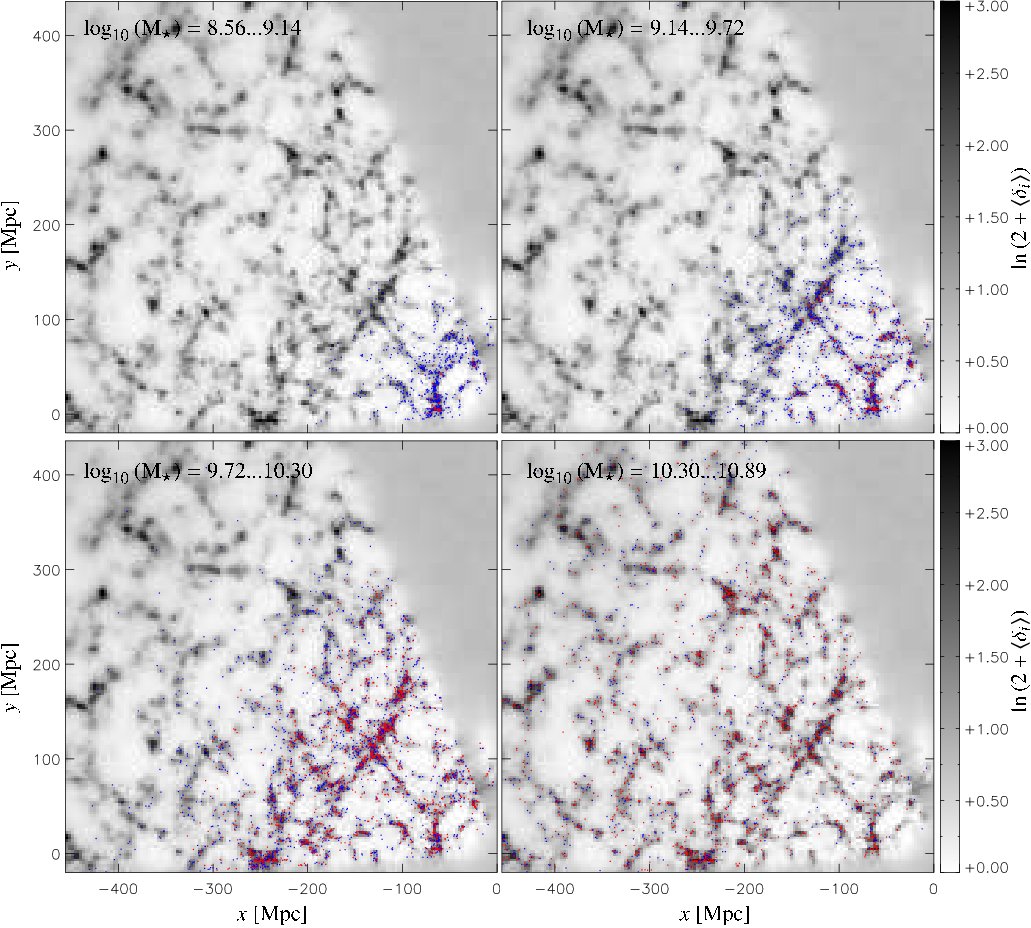}}
\caption{SDSS galaxies overplotted on the ensemble mean density field. The blue and red dots denote blue and red galaxies respectively, and the different panels depict galaxies in different stellar mass \(\rm{M}_{\star}\) bins.}
\label{fig:gal_dens}
\end{figure*}

\begin{figure*}
\centering{\includegraphics[width=1.0\textwidth,clip=true]{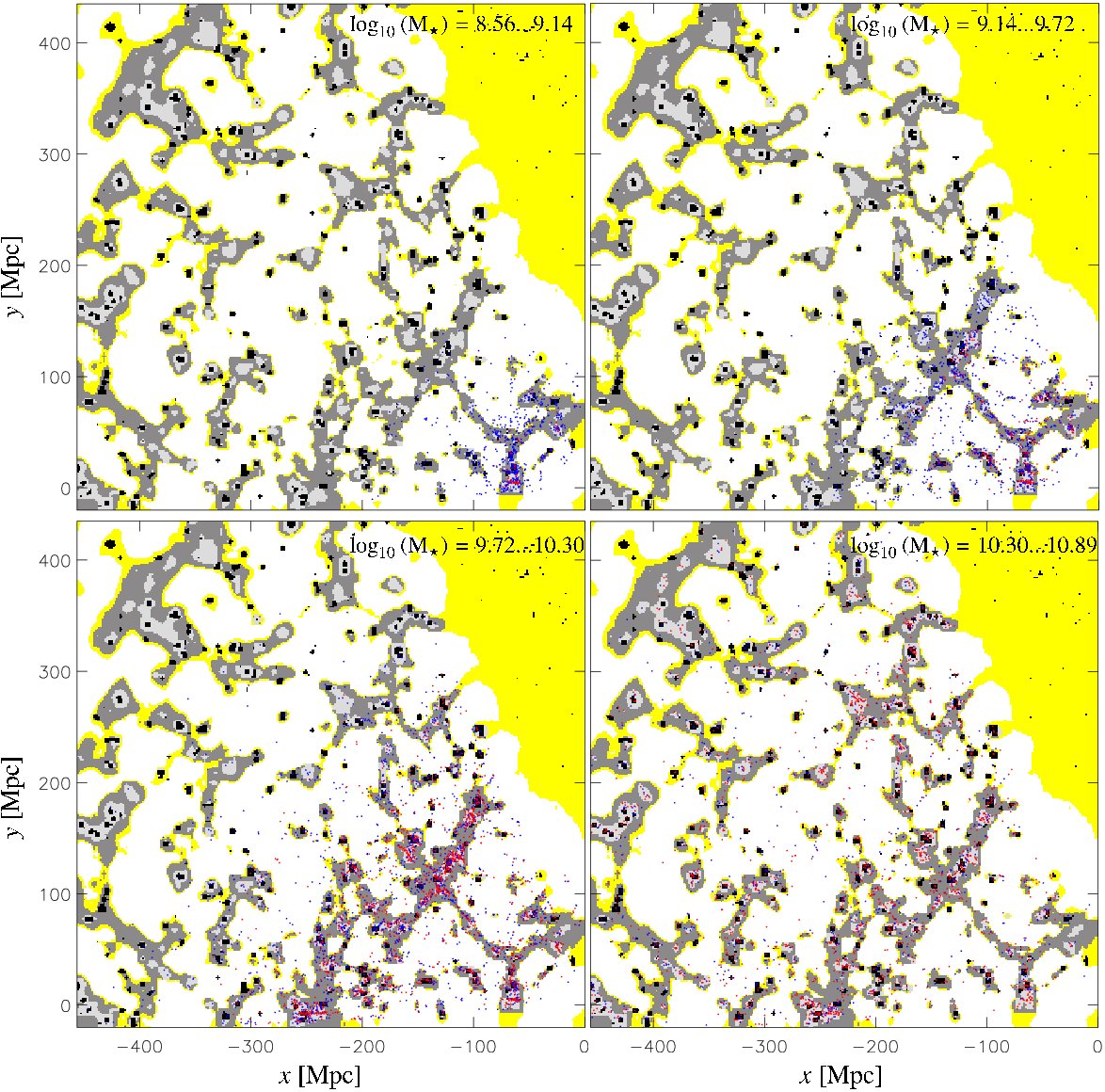}}
\caption{Same as figure \ref{fig:gal_dens}, but here the galaxies are overplotted on a structure type map as defined in section \ref{web_classification_data}. The color coding denotes the  web type: halo (black), filament (light grey), sheet (dark grey) and void (white). Regions, which are marked as undecided according to our criteria, equation (\ref{eqn:web_type_map}) with \(O_{th}=1.55\), are colored yellow.}
\label{fig:gal_web_type}
\end{figure*}

Our results are shown in figures  \ref{fig:gal_dens} and \ref{fig:gal_web_type} where we plot the galaxies
in our sample with different  stellar masses and $g-r$ colors, on top
of a slice through the ensemble mean density field. In each figure the
four panels  correspond to four  $M_\star$ intervals as  indicated. The
galaxies  falling into  a  given  $M_\star$ range  are  plotted in  the
corresponding  panel, with  red  (blue) galaxies  being  shown as  red
(blue) dots. Here we classify  each galaxy into red or blue population
using  its  $g-r$  color  and  the  luminosity-dependent  divider  as
described  in \citet{LI2006}  (see  their  Eq.  7  and  Table 4).   The
observer  on Earth is  at the  bottom right-hand  corner of  the slice
where $x=0$ and $y=0$ Mpc.  The density field with $z=302.16\pm 4.5$ Mpc is
projected onto  the $x-y$  plane and is  repeated in every  panel.  In
figure  \ref{fig:gal_dens}  the  background  density   field  is  coded  by  the  mean
overdensity, $\ln(2+\langle \delta_i \rangle)$,  averaged for each pixel  over the $z$
range probed  and the 40,000  Hamiltonian samples. In figure \ref{fig:gal_web_type} we present a structure type map as defined in equation (\ref{eqn:web_type_map}) by choosing an odds threshold of \(O_{th}=1.55\) and \(\lambda_{th}=1.0\). Each 
pixel of this map is color-coded  by the  web type  which is  determined  by our
classification  algorithm   described  above,  with   types  of  halo,
filament, sheet and void being plotted in black, light grey, dark grey
and white respectively.

Qualitatively the galaxies plotted  in these figures appear to closely
trace the  underlying large-scale  structure.  This is  not surprising
because,  by  construction,  the  latter  is  reconstructed  from  the
former. However, careful comparison of the different panels reveals a
number of interesting trends.  First, there exists a clear correlation
between galaxy mass and the large-scale environment, regardless of how
the environment is quantified. More massive galaxies tend to reside in
regions with  higher densities and more halo-like  structures.  At the
highest  masses, almost all  galaxies are  confined within  regions of
high  densities, or  those of  halo  and filament  types. As  $M_\star$
decreases,   more   and  more   galaxies   are   found  in   void-like
regions. Second, at fixed stellar  mass, galaxy color also appears to
be correlated  with large-scale  environment.  Red galaxies  trace the
density field  more closely  than blue galaxies.   At all  masses, the
distribution of  blue galaxies is  more extended across  the different
types of structures. At low  masses, the blue population dominates the
galaxies in void-like environment.

These trends are consistent with  recent similar studies by \citet{LEE2008} and \citet{LEELI2008},  which were based on  much shallower galaxy samples (thus  smaller volume), and also with  the clustering analyses
of  \citet{LI2006}.  More work  is needed  in order  to  have more
quantitative  characterization  of  the relationships  between  galaxy
properties  and the  large-scale environment,  and thus  more powerful
constrains on galaxy formation models. These results, in turn, can be fed back to the large scale structure inference and help to improve our cosmographical description of the Universe.

\section{Summary and Conclusion}
\label{SUMMARYANDCONCLUSION}
In this work we present the first application of the non-linear, non-Gaussian Bayesian large scale structure inference algorithm HADES to SDSS DR7 data.

HADES is a numerically efficient implementation of a Hamiltonian
Markov Chain sampler, which performs sampling in extremely high
parameter spaces usually consisting of \(\sim 10^7\) or more free
parameters. In particular, HADES explores the lognormal Poissonian
density posterior, which permits precision recovery of poorly sampled
objects and density field inference deep into the non-linear regime \citep[][]{JASCHE2009}.

The large scale structure inference was conducted on a cubic
equidistant grid with sidelength of \(750\) Mpc consisting of
\(256^3\) voxels, yielding a grid resolution of about \(3\) Mpc.
The large scale structure inference procedure correctly accounts for
the survey geometry, completeness and radial selection effects as well
as for the correct treatment of Poissonian noise.
The analysis yielded  about 3 TB of valuable scientific information in the
form of full three dimensional density samples of the lognormal
Poissonian density posterior. This set of density samples is thus a sampled
representation of the full non-Gaussian density posterior distribution
and therefore encodes all observational systematics and statistical
uncertainties. Hence, all uncertainties and systematics can
seamlessly be propagated to any finally inferred quantity, by simply
applying the according inference procedure to the set of
samples. In this fashion, the results permit us to make precise and
quantitative statements about the large scale density field and any
derived quantity.

We stress that our Hamiltonian samples are not the result of a filtering procedure.
A filter generally suppresses the power of the signal in low signal-to-noise
regions and therefore does not yield a physical meaningful density,
since it lacks power in poorly or unobserved regions.
However, each Hamiltonian density sample represents a complete physical matter field realization
conditional on the observations, in the sense that it possesses correct
physical power throughout the entire volume. Already visual inspection of these
density samples shows a homogeneous distribution of power throughout
the entire volume. This fact was emphasized by the demonstration of
power-spectra measured from these density samples, which show no sign
of being affected by lack of power or artificial mode coupling nor do
they show any sign of being affected by an adaptive smoothing kernel
as would be expected for filter applications.
It should be noted that this fact marks the crucial difference of our method to previous filter based density estimation procedures.

In section \ref{MEANANDVARIANCE}, we estimated the ensemble mean and the according variance from the \(40000\) density samples.
The estimated ensemble mean nicely depicts the cosmic web consisting
of filaments, voids and clusters extracted from the SDSS data. It is
clear, that the ensemble mean represents the mean estimated from the
lognormal Poissonian posterior conditional on the SDSS
data. Therefore, it encodes the observational uncertainties and
systematics. This can be seen by the fact, that the ensemble mean
approaches cosmic mean density in poorly or not observed
regions. Further, we plotted the according variance, which demonstrates
that the non-Gaussian behavior and structure of the Poissonian shot
noise was correctly taken into account in our analysis. Especially,
the expected correlation between high mean density and high variance
regions was clearly visible. We also estimated the cumulative
probabilities for the density amplitude at each volume element, and
demonstrated that the recovered density fields truly cover the broad
range from linear to non-linear density amplitudes.

To characterize the environment of our galaxy sample, but also to demonstrate the advantages of the Hamiltonian samples, we performed an example cosmic web
type classification in section
\ref{web_classification_data}. In particular, we followed the dynamical cosmic
web classification approach of \citet{HAHN2007} with the extensions proposed
by \citet{FORERO2009}. This procedure involves the calculation of the cosmic
deformation tensor and its eigenvalues. We demonstrated that this
procedure can easily be applied to the set of samples, since they
represent full physical matter field realizations. As a byproduct of this procedure we estimated the
ensemble mean for the three eigenvalues of the cosmic deformation tensor. Further, we classified the individual volume
elements as one of the four different web types void, sheet, filament
and halo. The classification into four discrete web types enabled us
to explicitely estimate the cosmic web posterior,
which provides the probability of finding a specific web type at a
given point in the volume conditional on the SDSS data. This result is
especially appealing from a Bayesian point of view, since it
emphasizes the fact, that the result of a Bayesian method is a
complete probability distribution rather than just a single
estimate. Here we saw, that especially voids are a sensitive measure
for the cosmic web.
Of course, it is possible to repeat the cosmic web classification in a
similar manner with any other classification procedure.

In the following section \ref{GALAXY_PROPERTIES}, we presented a
preliminary examination of the correlation between the large-scale
environment and physical properties of galaxies. In particular, we
considered the stellar mass and $g-r$ color of galaxies in relation to
the density contrast \(\delta\). A qualitative analysis revealed that
there exist correlation between these galaxy properties and the large
scale structure. In particular, massive galaxies are more likely to be
found in massive structures, while low mass galaxies reside in void
like structures. The plots also demonstrate the different clustering
behavior of red and blue galaxies. Also note, that these observed
trends are consistent with previous works
\citep[][]{LEE2008,LEELI2008,LI2006}. However, more work is required
in order to provide quantitative statements. This will be done in a
forthcoming publication.

The results presented in this work will be valuable for many
subsequent scientific analyses of the dependence of galaxy properties on their cosmic environment. Herefore, particularly the Hamiltonian
samples allow for a more intuitive handling of observational data,
since they can be understood as full matter field realizations or
different multiverses consistent with our data of the Universe we live in. Beside
providing quantitative characterizations of the large scale
structure, the results also give us an intuitive understanding of the three dimensional matter distribution
in our cosmic neighborhood. We intend to make our data publically available to the community.

Future applications will also take into account non-linear bias models and peculiar velocity sampling procedures, to provide even more accurate density analyses.

We hope that this work demonstrates the potential of Bayesian large
scale structure inference and its contribution to current and future
precision analyses of our Universe.

\section*{Acknowledgments}
We would like to thank Ofer Lahav and Benjamin D. Wandelt for suggesting us to use the lognormal Poissonian posterior for large scale structure inference. We also thank Simon D. M. White for encouraging discussions. Particular thanks also to Rainer Moll and Bj\"{o}rn Malte Sch\"{a}fer for usefull discussions and support with many valuable numerical gadgets. The authors also thank Benton R. Metcalf for many interesting discussions and comments on this project and Andreas Faltenbacher for suggesting us to estimate the cosmic deformation tensor. Special thanks also to Mar\'{i}a \'{A}ngeles Bazarra Castro for helpful assistance during the course of this project.
Further, we thank the "Transregional Collaborative Research Centre TRR 33 - The Dark Universe" for the support of this work.

Funding for the SDSS and SDSS-II has been provided by the Alfred P. Sloan Foundation, the Participating Institutions, the National Science Foundation, the U.S. Department of Energy, the National Aeronautics and Space Administration, the Japanese Monbukagakusho, the Max Planck Society, and the Higher Education Funding Council for England. The SDSS Web Site is http://www.sdss.org/.

The SDSS is managed by the Astrophysical Research Consortium for the Participating Institutions. The Participating Institutions are the American Museum of Natural History, Astrophysical Institute Potsdam, University of Basel, University of Cambridge, Case Western Reserve University, University of Chicago, Drexel University, Fermilab, the Institute for Advanced Study, the Japan Participation Group, Johns Hopkins University, the Joint Institute for Nuclear Astrophysics, the Kavli Institute for Particle Astrophysics and Cosmology, the Korean Scientist Group, the Chinese Academy of Sciences (LAMOST), Los Alamos National Laboratory, the Max-Planck-Institute for Astronomy (MPIA), the Max-Planck-Institute for Astrophysics (MPA), New Mexico State University, Ohio State University, University of Pittsburgh, University of Portsmouth, Princeton University, the United States Naval Observatory, and the University of Washington.

\bibliography{paper}

\begin{thebibliography}{}

\bibitem[\protect\citeauthoryear{{Abazajian} et~al.,}{{Abazajian}
  et~al.}{2009}]{SDSS7}
{Abazajian} K.~N.,  et~al., 2009, \apjs, 182, 543

\bibitem[\protect\citeauthoryear{Alcock \& Paczy{\'{n}}ski}{Alcock \&
  Paczy{\'{n}}ski}{1979}]{AP-79}
Alcock C.,  Paczy{\'{n}}ski B.,  1979, Nature, 281, 358

\bibitem[\protect\citeauthoryear{{Arag{\'o}n-Calvo}, {Jones}, {van de Weygaert}
  \& {van der Hulst}}{{Arag{\'o}n-Calvo} et~al.}{2007}]{ARAGON2007}
{Arag{\'o}n-Calvo} M.~A.,  {Jones} B.~J.~T.,  {van de Weygaert} R.,    {van der
  Hulst} J.~M.,  2007, \aap, 474, 315

\bibitem[\protect\citeauthoryear{{Ballinger}, {Peacock} \&
  {Heavens}}{{Ballinger} et~al.}{1996}]{1996MNRAS.282..877B}
{Ballinger} W.~E.,  {Peacock} J.~A.,    {Heavens} A.~F.,  1996, \mnras, 282,
  877

\bibitem[\protect\citeauthoryear{{Bernardi}, {Nichol}, {Sheth}, {Miller} \&
  {Brinkmann}}{{Bernardi} et~al.}{2006}]{BERNARDI2006}
{Bernardi} M.,  {Nichol} R.~C.,  {Sheth} R.~K.,  {Miller} C.~J.,    {Brinkmann}
  J.,  2006, \aj, 131, 1288

\bibitem[\protect\citeauthoryear{{Bistolas} \& {Hoffman}}{{Bistolas} \&
  {Hoffman}}{1998}]{BISTOLAS1998}
{Bistolas} V.,  {Hoffman} Y.,  1998, \apj, 492, 439

\bibitem[\protect\citeauthoryear{{Blanton}, {Brinkmann}, {Csabai}, {Doi},
  {Eisenstein}, {Fukugita}, {Gunn}, {Hogg} \& {Schlegel}}{{Blanton}
  et~al.}{2003}]{BLANTON2003}
{Blanton} M.~R.,  {Brinkmann} J.,  {Csabai} I.,  {Doi} M.,  {Eisenstein} D.,
  {Fukugita} M.,  {Gunn} J.~E.,  {Hogg} D.~W.,    {Schlegel} D.~J.,  2003, \aj,
  125, 2348

\bibitem[\protect\citeauthoryear{{Blanton}, {Eisenstein}, {Hogg}, {Schlegel} \&
  {Brinkmann}}{{Blanton} et~al.}{2005}]{BLANTON2005}
{Blanton} M.~R.,  {Eisenstein} D.,  {Hogg} D.~W.,  {Schlegel} D.~J.,
  {Brinkmann} J.,  2005, \apj, 629, 143

\bibitem[\protect\citeauthoryear{{Blanton}, {Lin}, {Lupton}, {Maley}, {Young},
  {Zehavi} \& {Loveday}}{{Blanton} et~al.}{2003}]{BLANTON2003A}
{Blanton} M.~R.,  {Lin} H.,  {Lupton} R.~H.,  {Maley} F.~M.,  {Young} N.,
  {Zehavi} I.,    {Loveday} J.,  2003, \aj, 125, 2276

\bibitem[\protect\citeauthoryear{{Blanton} \& {Roweis}}{{Blanton} \&
  {Roweis}}{2007}]{BLANTON2007}
{Blanton} M.~R.,  {Roweis} S.,  2007, \aj, 133, 734

\bibitem[\protect\citeauthoryear{{Cabr{\'e}} \& {Gazta{\~n}aga}}{{Cabr{\'e}} \&
  {Gazta{\~n}aga}}{2009}]{CABRE2009}
{Cabr{\'e}} A.,  {Gazta{\~n}aga} E.,  2009, \mnras, 396, 1119

\bibitem[\protect\citeauthoryear{{Choi}, {Park} \& {Vogeley}}{{Choi}
  et~al.}{2007}]{CHOI2007}
{Choi} Y.,  {Park} C.,    {Vogeley} M.~S.,  2007, \apj, 658, 884

\bibitem[\protect\citeauthoryear{{Colberg} et~al.,}{{Colberg}
  et~al.}{2008}]{COLBERG2008A}
{Colberg} J.~M.,  et~al., 2008, \mnras, 387, 933

\bibitem[\protect\citeauthoryear{{Colberg}, {Sheth}, {Diaferio}, {Gao} \&
  {Yoshida}}{{Colberg} et~al.}{2005}]{COLBERG2005}
{Colberg} J.~M.,  {Sheth} R.~K.,  {Diaferio} A.,  {Gao} L.,    {Yoshida} N.,
  2005, \mnras, 360, 216

\bibitem[\protect\citeauthoryear{{Coles} \& {Jones}}{{Coles} \&
  {Jones}}{1991}]{COLES1991}
{Coles} P.,  {Jones} B.,  1991, \mnras, 248, 1

\bibitem[\protect\citeauthoryear{Cowles \& Carlin}{Cowles \&
  Carlin}{1996}]{COWLES1996}
Cowles M.~K.,  Carlin B.~P.,  1996, Journal of the American Statistical
  Association, 91, 883

\bibitem[\protect\citeauthoryear{{Davis} \& {Peebles}}{{Davis} \&
  {Peebles}}{1983}]{DAVISPEEBLES1983}
{Davis} M.,  {Peebles} P.~J.~E.,  1983, \apj, 267, 465

\bibitem[\protect\citeauthoryear{{D'Mellow} \& {Taylor}}{{D'Mellow} \&
  {Taylor}}{2000}]{DMELLOW2000}
{D'Mellow} K.~J.,  {Taylor} A.~N.,  2000, in {S.~Courteau \& J.~Willick} ed.,
  Cosmic Flows Workshop Vol.~201 of Astronomical Society of the Pacific
  Conference Series, {Generalising the Inverse Redshift-Space Operator:
  Vorticity in Redshift-Space}.
p.~298

\bibitem[\protect\citeauthoryear{{Dressler}}{{Dressler}}{1980}]{DRESSLER1980}
{Dressler} A.,  1980, \apj, 236, 351

\bibitem[\protect\citeauthoryear{{Duane}, {Kennedy}, {Pendleton} \&
  {Roweth}}{{Duane} et~al.}{1987}]{DUANE1987}
{Duane} S.,  {Kennedy} A.~D.,  {Pendleton} B.~J.,    {Roweth} D.,  1987,
  Physics Letters B, 195, 216

\bibitem[\protect\citeauthoryear{{Dunkley}, {Bucher}, {Ferreira}, {Moodley} \&
  {Skordis}}{{Dunkley} et~al.}{2005}]{DUNKLEY2005}
{Dunkley} J.,  {Bucher} M.,  {Ferreira} P.~G.,  {Moodley} K.,    {Skordis} C.,
  2005, \mnras, 356, 925

\bibitem[\protect\citeauthoryear{{Ebeling} \& {Wiedenmann}}{{Ebeling} \&
  {Wiedenmann}}{1993}]{1993PhRvE..47..704E}
{Ebeling} H.,  {Wiedenmann} G.,  1993, \pre, 47, 704

\bibitem[\protect\citeauthoryear{{Eisenstein} \& {Hu}}{{Eisenstein} \&
  {Hu}}{1998}]{1998ApJ...496..605E}
{Eisenstein} D.~J.,  {Hu} W.,  1998, \apj, 496, 605

\bibitem[\protect\citeauthoryear{{Eisenstein} \& {Hu}}{{Eisenstein} \&
  {Hu}}{1999}]{1999ApJ...511....5E}
{Eisenstein} D.~J.,  {Hu} W.,  1999, \apj, 511, 5

\bibitem[\protect\citeauthoryear{{En{\ss}lin}, {Frommert} \&
  {Kitaura}}{{En{\ss}lin} et~al.}{2008}]{ENSSLIN2008}
{En{\ss}lin} T.~A.,  {Frommert} M.,    {Kitaura} F.~S.,  2008, ArXiv e-prints

\bibitem[\protect\citeauthoryear{{Erdo{\u g}du} et~al.,}{{Erdo{\u g}du}
  et~al.}{2004}]{ERDOGDU2004}
{Erdo{\u g}du} P.,  et~al., 2004, \mnras, 352, 939

\bibitem[\protect\citeauthoryear{{Erdo{\u g}du}, {Lahav}, {Huchra} \& {et
  al.}}{{Erdo{\u g}du} et~al.}{2006}]{2006MNRAS.373...45E}
{Erdo{\u g}du} P.,  {Lahav} O.,  {Huchra} J.,    {et al.} 2006, \mnras, 373, 45

\bibitem[\protect\citeauthoryear{{Fisher}, {Lahav}, {Hoffman}, {Lynden-Bell} \&
  {Zaroubi}}{{Fisher} et~al.}{1995}]{1995MNRAS.272..885F}
{Fisher} K.~B.,  {Lahav} O.,  {Hoffman} Y.,  {Lynden-Bell} D.,    {Zaroubi} S.,
   1995, \mnras, 272, 885

\bibitem[\protect\citeauthoryear{{Forero-Romero}, {Hoffman}, {Gottl{\"o}ber},
  {Klypin} \& {Yepes}}{{Forero-Romero} et~al.}{2009}]{FORERO2009}
{Forero-Romero} J.~E.,  {Hoffman} Y.,  {Gottl{\"o}ber} S.,  {Klypin} A.,
  {Yepes} G.,  2009, \mnras, 396, 1815

\bibitem[\protect\citeauthoryear{{Frommert}, {En{\ss}lin} \&
  {Kitaura}}{{Frommert} et~al.}{2008}]{FROMMERT2008}
{Frommert} M.,  {En{\ss}lin} T.~A.,    {Kitaura} F.~S.,  2008, \mnras, 391,
  1315

\bibitem[\protect\citeauthoryear{{Gaztanaga} \& {Yokoyama}}{{Gaztanaga} \&
  {Yokoyama}}{1993}]{GAZTANAGA1993}
{Gaztanaga} E.,  {Yokoyama} J.,  1993, \apj, 403, 450

\bibitem[\protect\citeauthoryear{Gelman \& Rubin}{Gelman \&
  Rubin}{1992}]{GELMAN1992}
Gelman A.,  Rubin D.,  1992, Statistical Science, 7, 457

\bibitem[\protect\citeauthoryear{Geweke}{Geweke}{1992}]{GEWEKE1992}
Geweke J., , 1992, Evaluating the Accuracy of Sampling-Based Approaches to the
  Calculation of Posterior Moments

\bibitem[\protect\citeauthoryear{{G{\'o}mez} et~al.,}{{G{\'o}mez}
  et~al.}{2003}]{GOMEZ2003}
{G{\'o}mez} P.~L.,  et~al., 2003, \apj, 584, 210

\bibitem[\protect\citeauthoryear{{Goto}, {Yamauchi}, {Fujita}, {Okamura},
  {Sekiguchi}, {Smail}, {Bernardi} \& {Gomez}}{{Goto} et~al.}{2003}]{GOTO2003}
{Goto} T.,  {Yamauchi} C.,  {Fujita} Y.,  {Okamura} S.,  {Sekiguchi} M.,
  {Smail} I.,  {Bernardi} M.,    {Gomez} P.~L.,  2003, \mnras, 346, 601

\bibitem[\protect\citeauthoryear{{Hahn}, {Porciani}, {Carollo} \&
  {Dekel}}{{Hahn} et~al.}{2007}]{HAHN2007}
{Hahn} O.,  {Porciani} C.,  {Carollo} C.~M.,    {Dekel} A.,  2007, \mnras, 375,
  489

\bibitem[\protect\citeauthoryear{{Hamann}, {Hannestad}, {Melchiorri} \&
  {Wong}}{{Hamann} et~al.}{2008}]{HAMANN2008}
{Hamann} J.,  {Hannestad} S.,  {Melchiorri} A.,    {Wong} Y.~Y.~Y.,  2008,
  Journal of Cosmology and Astro-Particle Physics, 7, 17

\bibitem[\protect\citeauthoryear{{Hamilton}}{{Hamilton}}{1998}]{hamilton-1998}
{Hamilton} A.~J.~S.,  1998, in {Hamilton} D.,  ed., The Evolving Universe
  Vol.~231 of Astrophysics and Space Science Library, {Linear Redshift
  Distortions: a Review}.
p.~185

\bibitem[\protect\citeauthoryear{{Hanson}}{{Hanson}}{2001}]{HANSON2001}
{Hanson} K.~M.,  2001, in {Sonka} M.,  {Hanson} K.~M.,  eds, Society of
  Photo-Optical Instrumentation Engineers (SPIE) Conference Series Vol.~4322 of
  Society of Photo-Optical Instrumentation Engineers (SPIE) Conference Series,
  {Markov chain Monte Carlo posterior sampling with the Hamiltonian method}.
pp 456--467

\bibitem[\protect\citeauthoryear{Heidelberger \& Welch}{Heidelberger \&
  Welch}{1981}]{HEIDELBERGER1981}
Heidelberger P.,  Welch P.~D.,  1981, Commun. ACM, 24, 233

\bibitem[\protect\citeauthoryear{Hockney \& Eastwood}{Hockney \&
  Eastwood}{1988}]{HOCKNEYEASTWOOD1988}
Hockney R.~W.,  Eastwood J.~W.,  1988, Computer simulation using particles.
Taylor \& Francis, Inc., Bristol, PA, USA

\bibitem[\protect\citeauthoryear{{Hoffman}}{{Hoffman}}{1994}]{HOFFMAN1994}
{Hoffman} Y.,  1994, in {Balkowski} C.,  {Kraan-Korteweg} R.~C.,  eds,
  Unveiling Large-Scale Structures Behind the Milky Way Vol.~67 of Astronomical
  Society of the Pacific Conference Series, {Wiener Reconstruction of the
  Large-Scale Structure in the Zone of Avoidance}.
p.~185

\bibitem[\protect\citeauthoryear{{Hubble}}{{Hubble}}{1934}]{HUBBLE1934}
{Hubble} E.,  1934, \apj, 79, 8

\bibitem[\protect\citeauthoryear{{Jasche} \& {Kitaura}}{{Jasche} \&
  {Kitaura}}{2009}]{JASCHE2009B}
{Jasche} J.,  {Kitaura} F.~S., , 2009, {Fast Hamiltonian Sampling for large
  scale structure inference}, submitted to MNRAS

\bibitem[\protect\citeauthoryear{{Jasche}, {Kitaura}, {Wandelt} \&
  {En{\ss}lin}}{{Jasche} et~al.}{2009}]{JASCHE2009}
{Jasche} J.,  {Kitaura} F.~S.,  {Wandelt} B.~D.,    {En{\ss}lin} T.~A., , 2009,
  {Bayesian power-spectrum inference for Large Scale Structure data}, submitted
  to MNRAS

\bibitem[\protect\citeauthoryear{{Jing} \& {B{\"o}rner}}{{Jing} \&
  {B{\"o}rner}}{2004}]{JINGBOERNER2004}
{Jing} Y.~P.,  {B{\"o}rner} G.,  2004, \apj, 617, 782

\bibitem[\protect\citeauthoryear{{Jing}, {Mo} \& {Boerner}}{{Jing}
  et~al.}{1998}]{JINGBOERNER1998}
{Jing} Y.~P.,  {Mo} H.~J.,    {Boerner} G.,  1998, \apj, 494, 1

\bibitem[\protect\citeauthoryear{{Kaiser}}{{Kaiser}}{1987}]{1987MNRAS.227....1%
K}
{Kaiser} N.,  1987, \mnras, 227, 1

\bibitem[\protect\citeauthoryear{{Kang}, {Jing}, {Mo} \& {B{\"o}rner}}{{Kang}
  et~al.}{2002}]{KANG2002}
{Kang} X.,  {Jing} Y.~P.,  {Mo} H.~J.,    {B{\"o}rner} G.,  2002, \mnras, 336,
  892

\bibitem[\protect\citeauthoryear{{Kayo}, {Taruya} \& {Suto}}{{Kayo}
  et~al.}{2001}]{KAYO2001}
{Kayo} I.,  {Taruya} A.,    {Suto} Y.,  2001, \apj, 561, 22

\bibitem[\protect\citeauthoryear{{Kitaura} \& {En{\ss}lin}}{{Kitaura} \&
  {En{\ss}lin}}{2008}]{Kitaura}
{Kitaura} F.~S.,  {En{\ss}lin} T.~A.,  2008, \mnras, 389, 497

\bibitem[\protect\citeauthoryear{{Kitaura}, {Jasche}, {Li}, {En{\ss}lin},
  {Metcalf}, {Wandelt}, {Lemson} \& {White}}{{Kitaura}
  et~al.}{2009}]{KITAURA2009}
{Kitaura} F.~S.,  {Jasche} J.,  {Li} C.,  {En{\ss}lin} T.~A.,  {Metcalf} R.~B.,
   {Wandelt} B.~D.,  {Lemson} G.,    {White} S.~D.~M.,  2009, ArXiv e-prints

\bibitem[\protect\citeauthoryear{{Kitaura}, {Jasche} \& {Metcalf}}{{Kitaura}
  et~al.}{2009}]{KITAURA2009B}
{Kitaura} F.~S.,  {Jasche} J.,    {Metcalf} R.~B.,  2009, ArXiv e-prints

\bibitem[\protect\citeauthoryear{{Klypin}, {Hoffman}, {Kravtsov} \&
  {Gottl{\"o}ber}}{{Klypin} et~al.}{2003}]{KLYPIN2003}
{Klypin} A.,  {Hoffman} Y.,  {Kravtsov} A.~V.,    {Gottl{\"o}ber} S.,  2003,
  \apj, 596, 19

\bibitem[\protect\citeauthoryear{{Kuehn} \& {Ryden}}{{Kuehn} \&
  {Ryden}}{2005}]{KUEHN2005}
{Kuehn} F.,  {Ryden} B.~S.,  2005, \apj, 634, 1032

\bibitem[\protect\citeauthoryear{{Lahav}}{{Lahav}}{1994}]{1994ASPC...67..171L}
{Lahav} O.,  1994, in {Balkowski} C.,  {Kraan-Korteweg} R.~C.,  eds, ASP Conf.
  Ser. 67: Unveiling Large-Scale Structures Behind the Milky Way {Wiener
  Reconstruction of All-Sky Spherical Harmonic Maps of the Large-Scale
  Structure}.
p.~171

\bibitem[\protect\citeauthoryear{{Lahav}, {Fisher}, {Hoffman}, {Scharf} \&
  {Zaroubi}}{{Lahav} et~al.}{1994}]{1994ApJ...423L..93L}
{Lahav} O.,  {Fisher} K.~B.,  {Hoffman} Y.,  {Scharf} C.~A.,    {Zaroubi} S.,
  1994, \apjl, 423, L93+

\bibitem[\protect\citeauthoryear{{Layzer}}{{Layzer}}{1956}]{LAYZER1956}
{Layzer} D.,  1956, \aj, 61, 383

\bibitem[\protect\citeauthoryear{{Lee} \& {Erdogdu}}{{Lee} \&
  {Erdogdu}}{2007}]{LEEANDERDOGDU2007}
{Lee} J.,  {Erdogdu} P.,  2007, \apj, 671, 1248

\bibitem[\protect\citeauthoryear{{Lee} \& {Lee}}{{Lee} \&
  {Lee}}{2008}]{LEE2008}
{Lee} J.,  {Lee} B.,  2008, \apj, 688, 78

\bibitem[\protect\citeauthoryear{{Lee} \& {Li}}{{Lee} \&
  {Li}}{2008}]{LEELI2008}
{Lee} J.,  {Li} C.,  2008, ArXiv e-prints

\bibitem[\protect\citeauthoryear{{Lemson} \& {Kauffmann}}{{Lemson} \&
  {Kauffmann}}{1999}]{LEMSON1999}
{Lemson} G.,  {Kauffmann} G.,  1999, \mnras, 302, 111

\bibitem[\protect\citeauthoryear{{Lewis} et~al.,}{{Lewis}
  et~al.}{2002}]{LEWIS2002}
{Lewis} I.,  et~al., 2002, \mnras, 334, 673

\bibitem[\protect\citeauthoryear{{Li}, {Kauffmann}, {Jing}, {White},
  {B{\"o}rner} \& {Cheng}}{{Li} et~al.}{2006}]{LI2006}
{Li} C.,  {Kauffmann} G.,  {Jing} Y.~P.,  {White} S.~D.~M.,  {B{\"o}rner} G.,
   {Cheng} F.~Z.,  2006, \mnras, 368, 21

\bibitem[\protect\citeauthoryear{{Libeskind}, {Yepes}, {Knebe}, {Gottloeber},
  {Hoffman} \& {Knollman}}{{Libeskind} et~al.}{2009}]{LIBESKIND2009}
{Libeskind} N.~I.,  {Yepes} G.,  {Knebe} A.,  {Gottloeber} S.,  {Hoffman} Y.,
   {Knollman} S.~R.,  2009, ArXiv e-prints

\bibitem[\protect\citeauthoryear{{Magira}, {Jing} \& {Suto}}{{Magira}
  et~al.}{2000}]{MAGIRA2000}
{Magira} H.,  {Jing} Y.~P.,    {Suto} Y.,  2000, \apj, 528, 30

\bibitem[\protect\citeauthoryear{{Mart{\'{\i}}nez} \& {Saar}}{{Mart{\'{\i}}nez}
  \& {Saar}}{2002}]{MARTINEZ2002}
{Mart{\'{\i}}nez} V.~J.,  {Saar} E.,  2002, {Statistics of the Galaxy
  Distribution}.
Chapman

\bibitem[\protect\citeauthoryear{{Martinez-Vaquero}, {Yepes}, {Hoffman},
  {Gottl{\"o}ber} \& {Sivan}}{{Martinez-Vaquero}
  et~al.}{2009}]{2009MNRAS.397.2070M}
{Martinez-Vaquero} L.~A.,  {Yepes} G.,  {Hoffman} Y.,  {Gottl{\"o}ber} S.,
  {Sivan} M.,  2009, \mnras, 397, 2070

\bibitem[\protect\citeauthoryear{{Matsubara} \& {Suto}}{{Matsubara} \&
  {Suto}}{1996}]{MATSUBARA1996}
{Matsubara} T.,  {Suto} Y.,  1996, \apjl, 470, L1

\bibitem[\protect\citeauthoryear{Neal}{Neal}{1993}]{NEAL1993}
Neal R.~M.,  1993, Technical Report CRG-TR-93-1, Probabilistic inference using
  Markov chain {M}onte {C}arlo methods.
University of Toronto

\bibitem[\protect\citeauthoryear{Neal}{Neal}{1996}]{NEAL1996}
Neal R.~M.,  1996, {Bayesian Learning for Neural Networks (Lecture Notes in
  Statistics)}, 1 edn.
Springer

\bibitem[\protect\citeauthoryear{{Novikov}, {Colombi} \& {Dor{\'e}}}{{Novikov}
  et~al.}{2006}]{NOVIKOV2006}
{Novikov} D.,  {Colombi} S.,    {Dor{\'e}} O.,  2006, \mnras, 366, 1201

\bibitem[\protect\citeauthoryear{{Park}, {Choi}, {Vogeley}, {Gott} \&
  {Blanton}}{{Park} et~al.}{2007}]{PARK2007}
{Park} C.,  {Choi} Y.,  {Vogeley} M.~S.,  {Gott} J.~R.~I.,    {Blanton} M.~R.,
  2007, \apj, 658, 898

\bibitem[\protect\citeauthoryear{{Peacock} \& {Dodds}}{{Peacock} \&
  {Dodds}}{1994}]{1994MNRAS.267.1020P}
{Peacock} J.~A.,  {Dodds} S.~J.,  1994, \mnras, 267, 1020

\bibitem[\protect\citeauthoryear{{Peacock} \& {Smith}}{{Peacock} \&
  {Smith}}{2000}]{PEACOCKSMITH2000}
{Peacock} J.~A.,  {Smith} R.~E.,  2000, \mnras, 318, 1144

\bibitem[\protect\citeauthoryear{{Peebles}}{{Peebles}}{1980}]{PEEBLES1980}
{Peebles} P.~J.~E.,  1980, {The large-scale structure of the universe}

\bibitem[\protect\citeauthoryear{{Percival} \& {White}}{{Percival} \&
  {White}}{2009}]{PERCIVAL2009}
{Percival} W.~J.,  {White} M.,  2009, \mnras, 393, 297

\bibitem[\protect\citeauthoryear{{Popowski}, {Weinberg}, {Ryden} \&
  {Osmer}}{{Popowski} et~al.}{1998}]{Popowski1998}
{Popowski} P.~A.,  {Weinberg} D.~H.,  {Ryden} B.~S.,    {Osmer} P.~S.,  1998,
  \apj, 498, 11

\bibitem[\protect\citeauthoryear{{Postman} \& {Geller}}{{Postman} \&
  {Geller}}{1984}]{POSTMAN1984}
{Postman} M.,  {Geller} M.~J.,  1984, \apj, 281, 95

\bibitem[\protect\citeauthoryear{Raftery \& Lewis}{Raftery \&
  Lewis}{1995}]{RAFTERY1995}
Raftery A.~E.,  Lewis S.~M.,  1995, in In Practical Markov Chain Monte Carlo
  (W.R. Gilks, D.J. Spiegelhalter and The number of iterations, convergence
  diagnostics and generic metropolis algorithms.
Chapman and Hall, pp 115--130

\bibitem[\protect\citeauthoryear{{Rojas}, {Vogeley}, {Hoyle} \&
  {Brinkmann}}{{Rojas} et~al.}{2005}]{ROJAS2005}
{Rojas} R.~R.,  {Vogeley} M.~S.,  {Hoyle} F.,    {Brinkmann} J.,  2005, \apj,
  624, 571

\bibitem[\protect\citeauthoryear{{Saunders} \& {Ballinger}}{{Saunders} \&
  {Ballinger}}{2000}]{SAUNDERS2000}
{Saunders} W.,  {Ballinger} W.~E.,  2000, in {Kraan-Korteweg} R.~C.,  {Henning}
  P.~A.,   {Andernach} H.,  eds, Mapping the Hidden Universe: The Universe
  behind the Mily Way - The Universe in HI Vol.~218 of Astronomical Society of
  the Pacific Conference Series, {Interpolation of Discretely-Sampled Density
  Fields}.
p.~181

\bibitem[\protect\citeauthoryear{{Saunders} et~al.,}{{Saunders}
  et~al.}{2000}]{SAUNDERS2000A}
{Saunders} W.,  et~al., 2000, in {R.~C.~Kraan-Korteweg, P.~A.~Henning, \&
  H.~Andernach} ed., Mapping the Hidden Universe: The Universe behind the Mily
  Way - The Universe in HI Vol.~218 of Astronomical Society of the Pacific
  Conference Series, {The IRAS View of the Local Universe}.
p.~141

\bibitem[\protect\citeauthoryear{{Scoccimarro}}{{Scoccimarro}}{2004}]{SCOCCIMA%
RRO2004}
{Scoccimarro} R.,  2004, \prd, 70, 083007

\bibitem[\protect\citeauthoryear{{Seljak}}{{Seljak}}{2000}]{SELJAK2000}
{Seljak} U.,  2000, \mnras, 318, 203

\bibitem[\protect\citeauthoryear{{Sheth}}{{Sheth}}{1995}]{SHETH1995}
{Sheth} R.~K.,  1995, \mnras, 277, 933

\bibitem[\protect\citeauthoryear{{Smith}, {Peacock}, {Jenkins}, {White},
  {Frenk}, {Pearce}, {Thomas}, {Efstathiou} \& {Couchman}}{{Smith}
  et~al.}{2003}]{SMITH2003}
{Smith} R.~E.,  {Peacock} J.~A.,  {Jenkins} A.,  {White} S.~D.~M.,  {Frenk}
  C.~S.,  {Pearce} F.~R.,  {Thomas} P.~A.,  {Efstathiou} G.,    {Couchman}
  H.~M.~P.,  2003, \mnras, 341, 1311

\bibitem[\protect\citeauthoryear{{Spergel} et~al.,}{{Spergel}
  et~al.}{2007}]{SPERGEL2007}
{Spergel} D.~N.,  et~al., 2007, \apjs, 170, 377

\bibitem[\protect\citeauthoryear{{Taylor} \& {Valentine}}{{Taylor} \&
  {Valentine}}{1999}]{TAYLOR1999}
{Taylor} A.,  {Valentine} H.,  1999, \mnras, 306, 491

\bibitem[\protect\citeauthoryear{Tegmark et~al.,}{Tegmark
  et~al.}{2004}]{sdss-tegmark}
Tegmark M.,  et~al., 2004, Phys. Rev. D, 69

\bibitem[\protect\citeauthoryear{{Tegmark} et~al.,}{{Tegmark}
  et~al.}{2006}]{TEGMARK2006}
{Tegmark} M.,  et~al., 2006, \prd, 74, 123507

\bibitem[\protect\citeauthoryear{{van de Weygaert} \& {Schaap}}{{van de
  Weygaert} \& {Schaap}}{2001}]{2001misk.conf..268V}
{van de Weygaert} R.,  {Schaap} W.,  2001, in {Banday} A.~J.,  {Zaroubi} S.,
  {Bartelmann} M.,  eds, Mining the Sky {Tessellation Reconstruction
  Techniques}.
p.~268

\bibitem[\protect\citeauthoryear{{Webster}, {Lahav} \& {Fisher}}{{Webster}
  et~al.}{1997}]{1997MNRAS.287..425W}
{Webster} M.,  {Lahav} O.,    {Fisher} K.,  1997, \mnras, 287, 425

\bibitem[\protect\citeauthoryear{{Whitmore}, {Gilmore} \& {Jones}}{{Whitmore}
  et~al.}{1993}]{WHITMORE1993}
{Whitmore} B.~C.,  {Gilmore} D.~M.,    {Jones} C.,  1993, \apj, 407, 489

\bibitem[\protect\citeauthoryear{{York} et~al.,}{{York}
  et~al.}{2000}]{YORK2000}
{York} D.~G.,  et~al., 2000, \aj, 120, 1579

\bibitem[\protect\citeauthoryear{{Zaninetti}}{{Zaninetti}}{1995}]{1995A&AS..10%
9...71Z}
{Zaninetti} L.,  1995, \aaps, 109, 71

\bibitem[\protect\citeauthoryear{{Zaroubi}, {Hoffman} \& {Dekel}}{{Zaroubi}
  et~al.}{1999}]{1999ApJ...520..413Z}
{Zaroubi} S.,  {Hoffman} Y.,    {Dekel} A.,  1999, \apj, 520, 413

\bibitem[\protect\citeauthoryear{{Zaroubi}, {Hoffman}, {Fisher} \&
  {Lahav}}{{Zaroubi} et~al.}{1995}]{1995ApJ...449..446Z}
{Zaroubi} S.,  {Hoffman} Y.,  {Fisher} K.~B.,    {Lahav} O.,  1995, \apj, 449,
  446

\end{thebibliography}
\bibliographystyle{mn2e}
%%% Local Variables: 
%%% mode: latex
%%% TeX-master: t
%%% End: 

\appendix

\bsp

\label{lastpage}

\end{document}